\documentclass[11pt]{article}
\usepackage{wrapfig}
\usepackage{epsfig}
\usepackage{graphicx, color}
\usepackage[breaklinks,colorlinks,urlcolor=blue,citecolor=blue,linkcolor=blue]{hyperref}
\usepackage{threeparttable}
\usepackage{amssymb,amsmath}

\makeatletter
\renewcommand\@biblabel[1]{\hspace{-\labelsep}}
\makeatother

\def\begref{\parindent=0pt\frenchspacing\parskip=0pt
\everypar={\hangindent=0.42in}}
\def\HI {H\thinspace{\sc i}}
\def\txs {TXS\thinspace2226{\tt -}184}
\def\pks {PKS\thinspace2322{\tt -}123}

\def\etal{{\it et al.}}
\def\deg{$^\circ$}
\def\nref{\parskip0pt\par\noindent\hangindent\parindent\hangafter1}

\begin{document}

\centerline{\Large A Next Generation Low Band Observatory: A Community}
\centerline{\Large Study Exploring Low Frequency Options for ngVLA}
\vspace{0.5cm}
\centerline{Greg Taylor, Jayce Dowell, Joe Malins (UNM), Tracy Clarke, Namir Kassim, Simona Giacintucci,}
\centerline{Brian Hicks, Jason Kooi, Wendy Peters, Emil Polisensky (NRL), Frank Schinzel \& Kevin Stovall (NRAO)}

\begin{abstract}

We present a community study exploring the low frequency (5 - 800 MHz) options and opportunities for the ngVLA project and its infrastructure.  We describe a Next Generation LOw Band Observatory (ngLOBO) that will provide access to the low frequency sky in a commensal fashion, operating independently from the ngVLA, but leveraging common infrastructure.  This approach provides continuous coverage through an aperture array (called ngLOBO-Low) below 150 MHz and by accessing the primary focus of the ngVLA antennas (called ngLOBO-High) above 150 MHz. ngLOBO preconditions include a) non-interference and b) low relative cost ($<$5\%) with respect to ngVLA.

ngLOBO has three primary scientific missions: (1) Radio Large Synoptic Survey Telescope (Radio-LSST): one naturally wide beam, commensal with ngVLA, will conduct a continuous synoptic survey of large swaths of the sky for both slow and fast transients; (2) This same commensal beam will provide complementary low frequency images of all ngVLA targets and their environment {\it when such data enhances their value}. (3) Independent beams from the ngLOBO-Low aperture array will conduct research in astrophysics, Earth science and space weather applications, engaging new communities and attracting independent resources. If ngVLA operates down to 2 GHz or lower, ngLOBO data will enhance ngVLA calibration and dynamic scheduling. Finally, non-variable field sources outside the ngVLA field of view can be harvested for serendipitous science, e.g. population studies for thermal and non-thermal continuum sources.

We will provide the benefit of our experience in building the Long Wavelength Array (LWA) stations including procuring land and running power and fiber in New Mexico. We also leverage our experience designing and operating the VLA Low-band Ionosphere and Transient Experiment (VLITE), a commensal prime-focus system currently acquiring $>$6000 hours of observing time annually from the VLA. VLITE is funded by DoD for ionospheric research and its expansion into ngLOBO is already of proven interest to distinct scientific communities and sponsors.

The ngVLA will be a superb, high frequency instrument; ngLOBO will allow it to participate in the worldwide renaissance in low frequency science as well.

\end{abstract}

\vfill
\eject

%\fcolorbox{gray}{orange}{\texttt{\textbackslash todo}}
%\fcolorbox{gray}{blue!40}{\texttt{\textbackslash todostyle}}
%\fcolorbox{gray}{yellow!40}{\texttt{\textbackslash todofigure}}
%\fcolorbox{gray}{purple!40}{\texttt{\textbackslash todolatex}}
%\fcolorbox{gray}{green!40}{\texttt{\textbackslash tododetail}}
%\fcolorbox{gray}{olive}{\texttt{\textbackslash todotable}}
%\fcolorbox{gray}{teal}{\texttt{\textbackslash todocitation}}
%
%\listoftodos
%
%\vfill
%\eject

\tableofcontents

\vfill
\eject

\vspace{-0.5cm}
\section{Introduction}
\vspace{-0.25cm}

We have been among the principals involved establishing and expanding low frequencies in New Mexico on the NRAO VLA and beyond. We have played key roles in the initial exploration and subsequent development of the 74 and 330 MHz feeds and receivers on the VLA (Kassim et al. 2007) as well as the recent expansion of the receivers to broadband systems compatible with the Karl G. Jansky VLA (Clarke et al. 2011). These systems have been in nearly continuous use as a regular part of the general observer VLA programs since the 330 MHz system was first installed in the 1980's. 
%\tododetail{TEC: Would like to quote here 4+P paper stats -- available somewhere?}

\begin{figure}[t!]
\begin{center}
\vspace{-1cm}
\includegraphics[width=5.5in,angle=0]{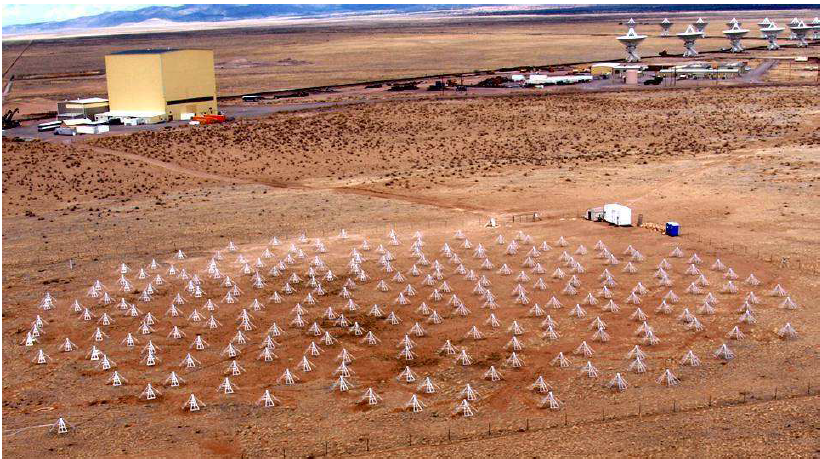}
\end{center}
\vspace{-0.5cm}
  \caption{
Aerial photograph of the LWA1 station, located near the center of
the Very Large Array on the Plains of San Augustin in central
New Mexico.
}
\label{lwa1}
\end{figure}

Beyond the VLA efforts, we have led the efforts to build and operate
the Long Wavelength Array\footnote[1]{\noindent see
  \url{http://lwa.unm.edu}} (LWA) which provides access to the low
frequency radio sky between 10 and 88 MHz to investigators through two
telescope sites located in New Mexico.  The first station (LWA1; see
Fig.~\ref{lwa1}) is co-located with the VLA, while the second New
Mexico station (LWA-SV; Cranmer et al. 2017) is located 70 km
NorthEast of LWA1 at the Sevilleta National Wildlife Refuge.  The LWA
is operated by the University of New Mexico for purposes of scientific
research at low frequencies.  The areas of research supported range
from ionospheric and space weather studies (Helmboldt et al. 2013,
2016), to meteors (Obenberger et al. 2014), to planetary
investigations (Clarke et al. 2014), to pulsars (Stovall et al. 2015),
and the nature of the first luminous bodies in the Universe (Kocz et
al. 2014; Taylor et al. 2012).  The LWA1 has two primary modes of
operation - (1) beamforming with between 1 and 4 beams, each 20 MHz
wide with dual polarization and two spectral tunings; (2) continuous
all-sky imaging in a narrow bandwidth (100 kHz).  The LWA-SV station
also supports both of these modes of operation though with a current
limitation of a single beam, and the added capability of a wideband
correlator mode.  Recently we also have begun to operate the two LWA
stations with the new 50-80 MHz modified J-pole (MJP) capability on
the VLA.  This combined instrument, called ELWA, provides spatial
resolution of 10 arc seconds and is a powerful array for low frequency
investigations.

Concurrently, we have developed the VLA Low-band Ionosphere and Transient Experiment\footnote[2]{\noindent see
\url{http://vlite.nrao.edu}} (VLITE), a purely commensal observing system on the NRAO VLA. The separate optical path of the prime-focus sub-GHz dipole feeds and the Cassegrain-focus GHz feeds provided an opportunity to expand the simultaneous frequency operation of the VLA through joint observations across both systems. The low-band receivers on 10 VLA antennas are outfitted with dedicated samplers and use spare fibers to transport the 320-384 MHz band to the VLITE correlator. The initial phase of VLITE uses a custom-designed real-time DiFX software correlator and operates at nearly 70\% wall time with over 6000 hours of VLA time recorded each year. VLITE data are used in real-time for ionospheric research and are transferred daily to NRL for processing in the astrophysics and transient pipelines. These pipelines provide automated radio frequency interference excision, calibration, imaging and self-calibration of data. VLITE scientific focus areas include the ionosphere, slow ($>$ 1 sec) transients, and astrophysics. VLITE is currently being expanded to 15 antennas, and will soon incorporate a parallel capability to search for fast ($<$ 1 sec), dispersed transients, including Fast Radio Bursts (FRBs) and Rotating Radio Transients (RRATs). VLITE has recently been enhanced to enable the production of data products to complement the upcoming NRAO 3 GHz VLA Sky Survey (VLASS).  VLITE data are already being used to provide serendipitous population studies of continuum sources across its large field of view. This will soon be expanded to spectroscopy, including OH Gigamasers and HI probing redshifts near z$\sim$4. ngLOBO applications in these areas are reviewed in the Appendix.

In the pre-ngVLA timeframe, we aim to grow the LWA and VLITE capabilities around the current VLA infrastructure. LWA science will evolve from mainly time-domain (single-dish) to imaging (aperture synthesis) science, through incorporation of VLA dishes equipped with MJP dipoles, and additional LWA stations. VLITE can expand to more VLA dishes and to include the full ($>$200 MHz) bandwidth of the VLA Low Band receivers. (At the time of this writing the original 10-antenna VLITE is being expanded to 15+ antennas.) LWA and VLITE can be unified into a single system, sharing a common backend, called the LOw Band Observatory (LOBO). 

In this white paper, we explore a longer term vision to grow a next generation (ngLOBO) system alongside the ngVLA. The low frequency (5 - 150 MHz) end, referred to as ngLOBO-Low consists of a dipole array located in close proximity to 50 of the ngVLA stations.  The high frequency (150 - 800 MHz) range is known as ngLOBO-High uses modest bandwidths ($\leq$ 1.5 GHz) and uncooled electronics coupled into the ngVLA dish infrastructure.  Both of these systems can be deployed at a fraction of the cost of ngVLA, and probably less than the cost of the fiber and power infrastructure that they would share with ngVLA. ngLOBO operates as a commensal system without interference with the ngVLA, and given its copious ``free'' observing time and naturally large fields-of-view delivers considerable added and independent scientific value. 

\vspace{-0.5cm}
\section{Existing and Planned Low Frequency Facilities}
\vspace{-0.25cm}

With advances in digital signal processing and calibration techniques, the entire field of low frequency radio astronomy has seen a renaissance. Most of this has also driven the developments for LWA and VLITE, and has created a boom in the construction of many low cost, low frequency radio observatories around the globe. In this section we provide a brief overview of the major existing facilities and their observing capabilities. 

\begin{itemize}
	\item {\bf OVRO-LWA/LEDA}: Located at the Owens Valley Radio Observatory, this array consists of 256 crossed dipoles of the LWA design operating between 40 and 88 MHz and extending to 2000 m baselines.  Correlation is provided by the Large-Aperture Experiment to Detect the Dark Ages (LEDA) correlator (Kocz et al. 2014) which correlates 60 MHz of bandwidth on timescales as short as 1 second.  The primary science drivers are slow ($>$1 second) transient sources which can be detected across the observable hemisphere and observations of the global 21 cm signal from the Dark Ages.
    \item {\bf MWA}: The Murchison Widefield Array in Western Australia is a dipole array that operates between 80 and 300 MHz. The MWA has 2048 dipoles arranged in 128 "tiles" that cover baselines up to 3 km.  The four main science themes are the EoR, galactic populations, radio transients and space weather.
    \item {\bf LOFAR}: The Low Frequency ARray (LOFAR) is located in Europe with the core stations and correlator situated in the Netherlands. Of the instruments described here, LOFAR is the most similar in resolution and theoretical imaging performance to ngLOBO. There are three types of LOFAR stations a) 18 core stations b) 18 remote stations, and currently 8 international stations with additional stations to be added in the future. Each LOFAR station consists of both low band (30 -80 MHz) and a high band (120-240 MHz) dipole arrays of various sizes and configurations. The baseline length of the stations ranges from 100 m to 1500 km. Significant differences from the 256-dipole LWA stations are that the core and remote stations stations employ only 48 dipoles at a given time. Furthermore,
the low band dipoles are resonant at 60 MHz and degrade in performance away from this frequency (van Haarlem et al. 2013). The individual LOFAR low band stations (except for the ``superterp'' which is a combination of 6 core stations) are much less sensitive than LWA stations. Key science drivers for LOFAR are the EoR, solar, transients, sky surveys, and cosmic magnetism.
    \item {\bf CHIME}: The Canadian Hydrogen Intensity Mapping Experiment consists of four large cylindrical reflectors feeding an array of radio receivers operating from 400-800 MHz.  This experiment is focused on measuring the baryon acoustic oscillation using redshifted 21 cm radiation from galaxies. Additionally, the array will be used for commensal fast radio transient searches. 
    \item {\bf VLA}: The current Very Large Array operates from 50 MHz to 50 GHz.  The feeds are a mix of offset cassegrain and prime focus (or near prime focus for the lowest frequency band).  The array elements are each 25 m in diameter and extend up to baselines of 35 km.  The science is primarily user driven with some major surveys.
    \item {\bf GMRT}:  The Giant Metre-Wave Radio Telescope in India operates between 150 and 1450 MHz using conventional prime focus feeds and 45-m diameter parabolic reflectors.  The distribution of the 30 GMRT antennas is in roughly a Y configuration and ranges over baselines up to 25 km.  The science is primarily user driven although there have been some surveys. The GMRT will soon be upgraded to a fully digital, broad-band system in much the same way that the legacy VLA was upgraded to the current JVLA.
    \item {\bf PAPER/HERA}: The Hydrogen Epoch of Recombination Array succeeds the Precision Array for Probing the Epoch of Reionization. HERA is a grid of 14 m non-tracking dishes packed into a hexagonal grid 300 m across.  Operating in the 100-200 MHz frequency range HERA is primarily concentrated on detecting the power spectrum of 21 cm radiation produced during the EOR.
    \item {\bf UTR-2}: The Ukranian T-shaped Radio telescope, second modification is an array of 2040 dipoles in two arms each containing 6 rows of elements.  The telescope operates between 8 and 32 MHz.  Primary science goals are solar and planetary astronomy.
    \item {\bf SKA-low}: The Square Kilometer Array project is currently under design so unlike the instruments described above it is still subject to modification.  As of mid-2017 the Low Frequency Aperture Array is planned to cover 50-200 MHz with a large number of dipoles of unknown design.  The configuration is not well defined but may extend out to baselines of $\sim$50 km.
\end{itemize}

\begin{figure}[t!]
\begin{center}
\vspace{-1cm}
\includegraphics[width=5.5in,angle=0]{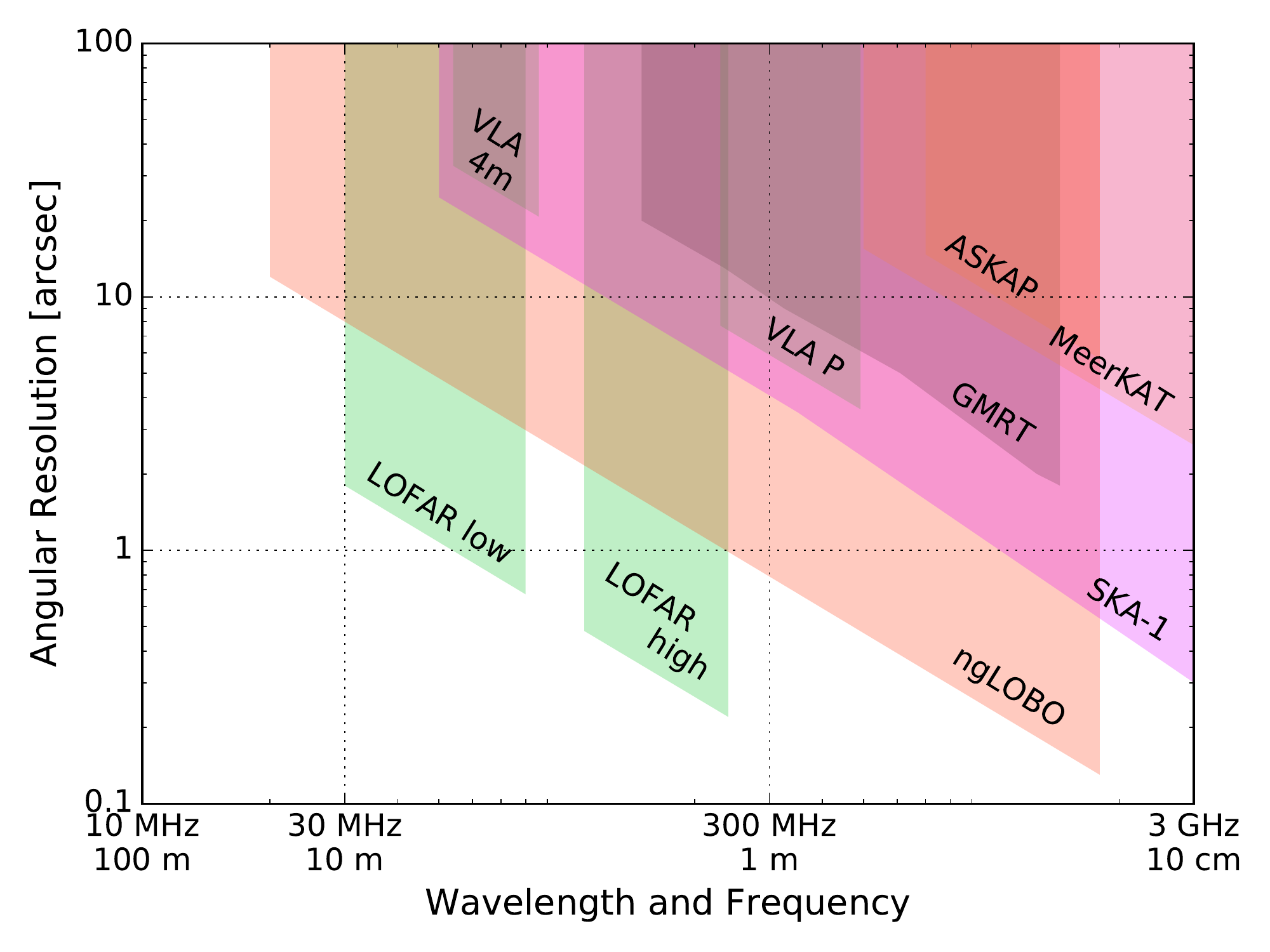}
\end{center}
\vspace{-0.5cm}
  \caption{
Angular resolution for a variety of low frequency arrays as a function of wavelength.
}
\label{angular}
\end{figure}

\begin{figure}[h!]
\begin{center}
\vspace{-1cm}
\includegraphics[width=5.5in,angle=0]{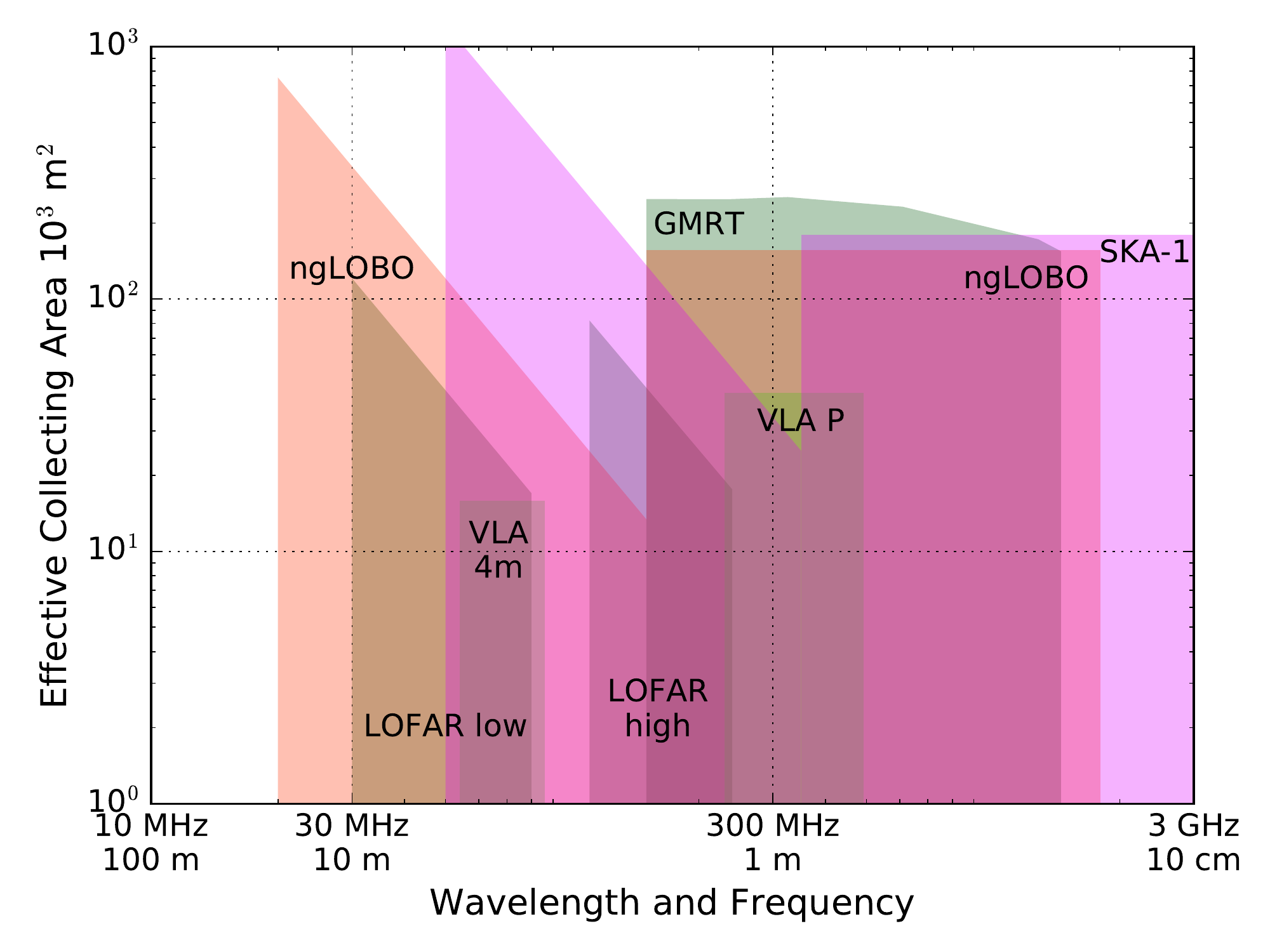}
\end{center}
\vspace{-0.5cm}
  \caption{
Collecting area for a variety of low frequency arrays as a function of wavelength.
}
\label{area}
\end{figure}

For purposes of comparison, 
in Fig.~\ref{angular} we show the frequency coverage and spatial resolution of all the aforementioned arrays as well as what ngLOBO would cover.  In Fig.~\ref{area} we show the collecting area for the arrays, noting that this plot does not take into account antenna/dipole efficiencies.

\vspace{-0.5cm}
\section{ngLOBO as a Radio LSST: The Frontier of Low Frequency Radio Transients}
\vspace{-0.25cm}

As highlighted by ASTRO2010, the dynamic radio sky has been poorly explored, mainly for technical reasons. The phase space of sub-GHz transients is especially ripe for exploration, owing to the confusion-limited sensitivity of past blind surveys and the non-thermal spectra of most expected source classes. As shown below, a significant improvement in sensitivity will certainly reveal expected populations, while a smattering of unidentified transients hints at larger, unknown populations. 

Transient and variable searches call for maximizing $\Omega \times t$, the product of the telescope field of view ($\Omega$) and observing time ($t$). Dish-based telescopes are normally inefficient, because time for lengthy blind surveying is generally unavailable, and $\Omega$ is not large at centimeter and shorter wavelengths. ngLOBO-High strikes a sweet spot by naturally inflating both $\Omega$ and $t$ at intermediate frequencies, offering transient monitoring with significant advantages over higher frequency ($>$ 1.5 GHz) searches.  It also has an advantage over aperture arrays because calibration is more robust.  Below 150 MHz, aperture arrays are unavoidable (e.g. LWA, LOFAR, MWA, SKA-Low), and naturally defines the transition to an aperture-array based ngLOBO-Low.

Both components of ngLOBO offer a natural extension to several source classes reviewed in the ngVLA Science Working Group 4 White paper (hereafter WGWP4: ``Time Domain,  Cosmology, Physics''). At lower frequencies there will be expected and serendipitous transients, notably steep spectrum coherent sources accessible to ngLOBO with no ngVLA counterparts. Table 1 lists a subset of expected and predicted source classes, a few of which are highlighted  below.

\subsection{Slow Transients and Incoherent Non-thermal Emission}

A natural mode of ngLOBO-High observations for slow ($>$ 1 sec),  transients, analogous to VLITE on the VLA, is a pipeline delivering calibrated continuum images and source catalogs. The wide-field, fixed FoV images will be generic, pathological fields aside (e.g. “A team” sources) and with time they will spread to cover a larger fraction, and depth, of the sky. With the cadence afforded revisiting popular ngVLA targets (e.g. M31, Galactic center, etc), many fields will be revisited often, and also observed by ngLOBO-Low. A radio analogy with LSST across the full ngLOBO frequency range follows naturally, providing a synoptic vision of the radio sky. 

An ngLOBO-based Radio LSST would benefit from an initial sky map, much as VLITE now depends on the VLSSr (Lane et al.\ 2014), WENSS (Rengelink et al.\ 1997), NVSS (Condon et al.\ 1998), and more recently, TGSS-ADR1 (Intema et al.\ 2017) and GLEAM (Hurley-Walker et al.\ 2017). These ``sky truth'' data provide a needed quiescent grid against which to register ngLOBO images for long duration variable sources, slow transients, or other changes.  

\subsubsection{Non-radio selected Slow Transients}

\begin{figure}[h!]
\begin{center}
\includegraphics[width=5.5in,angle=0]{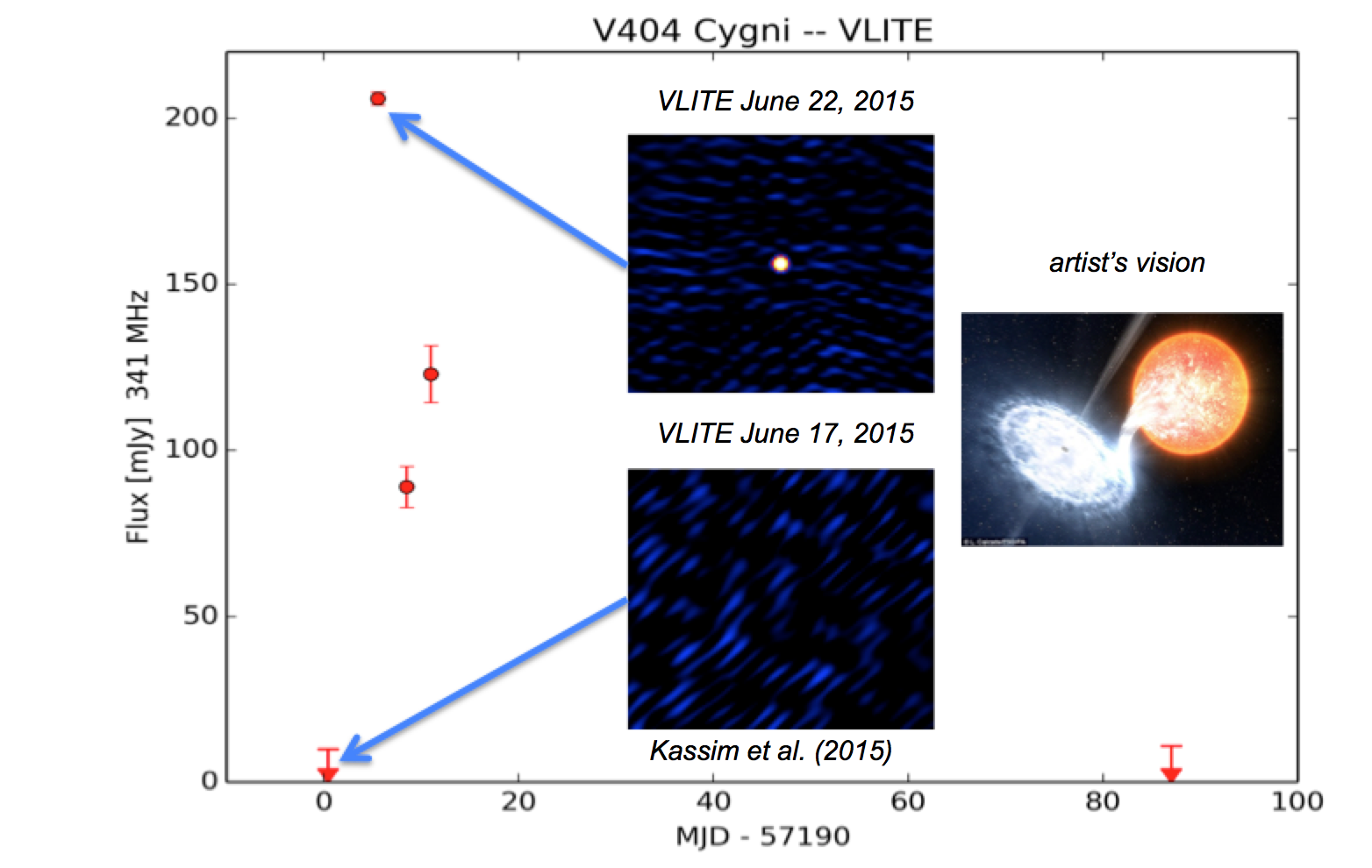}
\end{center}
\vspace{-0.5cm}
  \caption{
The LMXB V404 Cyg was detected in outburst in June 2015. The commensal VLITE system was used to trace the source from pre-outburst, through the burst to the post-burst quiescence. VLITE flux measurements for 5 epochs are shown with insets showing the VLITE image of the region pre-outburst (no detected source), during the outburst (V404 Cyg clearly detected) and an artists concept of the system.
}
\label{V404Cyg}
\end{figure}

\begin{figure}[t!]
\begin{center}
\includegraphics[width=5.5in,angle=0]{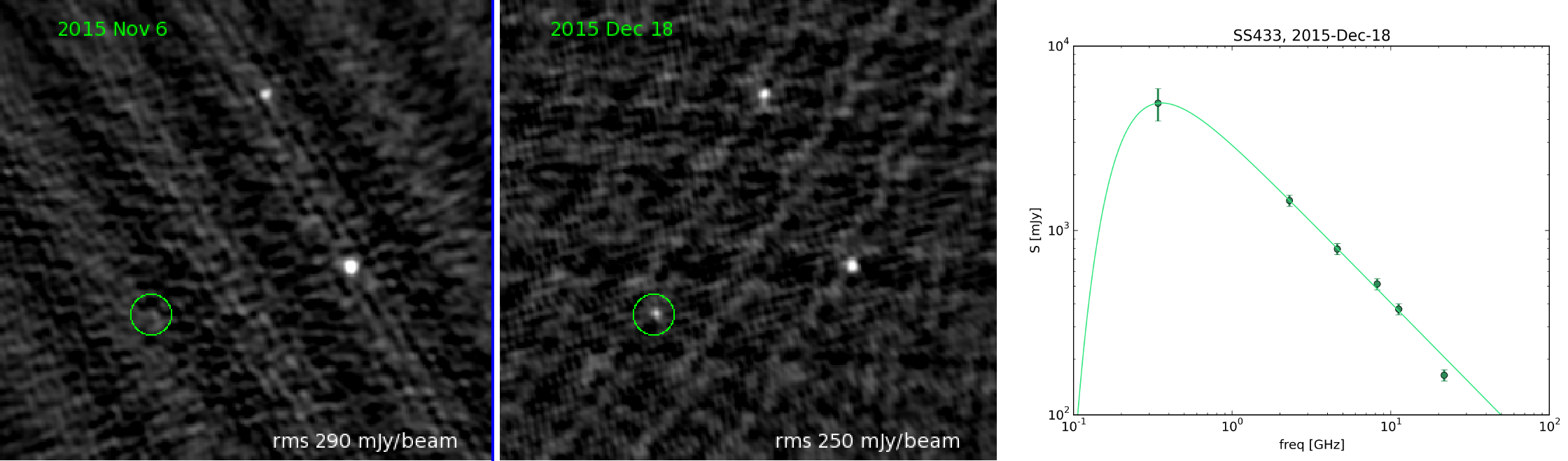}
\end{center}
\vspace{-0.5cm}
  \caption{
\textit{Left:} VLITE detection of outburst from microquasar SS433. \textit{Right:} Power-law spectrum with thermal HII absorption component fitted to outburst flux measurements from VLITE and RATAN-600. 
}
\label{SS433}
\end{figure}

To date, most low frequency transients have been discovered through targeted, follow-up observations to sources discovered at higher frequencies, typically in the optical (e.g. SNe, GMRT reference) and X-ray (e.g. X-ray binaries). Examples of VLITE detections of targeted sources include the LMXB V404 Cyg (Fig.~\ref{V404Cyg}) and the microquasar SS433 (Fig.~\ref{SS433}). Another recent example is the detection or radio emission from the flare star UV Ceti by the MWA (Lynch et al.\ 2017). 

\subsubsection{Radio-selected Slow Transients}

\begin{figure}[t!]
\begin{center}
\vspace{-1cm}
\includegraphics[width=5.in,angle=0]{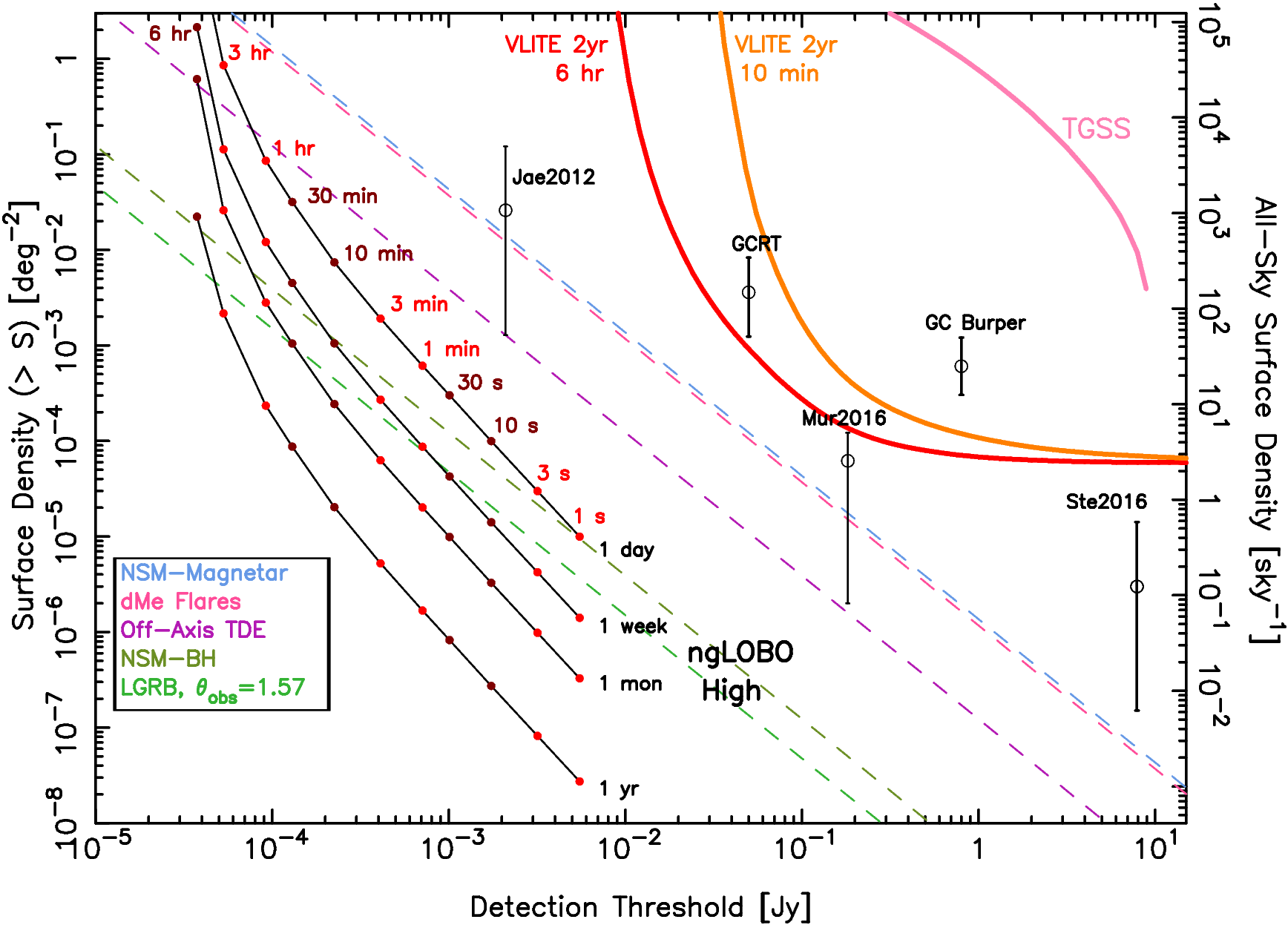}
\end{center}
\vspace{-0.5cm}
  \caption{Radio transient areal surface densities plotted against survey detection threshold for blind transient searches at sub-GHz frequencies. Points with $95\%$ confidence bars are bounds from surveys with transient detections. Diagonal dashed lines are estimated densities of the more common known transient source classes while the ``TGSS'' curve shows the density of quiescent 150 MHz sources in the TGSS survey. Red and orange lines are upper limits for 6 hour and 10 min timescale transients, respectively, after 2 years of VLITE operations. Black curves with points show conservative limits for transient timescales from 1~s to 6~hrs after 1 day, 1 week, 1 month and 1 year of ngLOBO-High operations.}  
\label{ngloboAREAL}
\end{figure}

More important is discovering new populations of $\textit {radio selected}$ transients through efficient, wide-field, low frequency observations. While a handful of sources have been detected, e.g. towards the Galactic center (Hyman et al. 2005) and elsewhere (Jaeger et al.\ 2012, Stewart et al.\ 2016, Murphy et al.\ 2017a), the numbers detected are modest. \textit{This is unsurprising given the limitations of current low frequency instruments for blind transient surveys.} Reasons include limited sensitivity, field-of-view, time-on-sky, or combinations thereof. This is demonstrated in Fig.~\ref{ngloboAREAL}, which illustrates the significant advances that are made from VLITE 
to ngLOBO, {\it detecting most classes of predicted low frequency transients}.

Fig.~\ref{ngloboAREAL} validates the claim that low frequency transients are a frontier right now. The known populations lying below current VLITE sensitivity are only slowly being uncovered and understood.  More exciting is the large, unprobed region of transient phase space lying between the capabilities of VLITE and ngLOBO. Low frequencies have a tremendous advantage because of the often non-thermal nature of the emission.

\begin{table}
\centering
\vspace{0.2cm}
\begin{threeparttable}
\caption{Transient source classes}\label{tabTRANS}
%\begin{center}
\begin{tabular}{l l l|r r r r r}
\hline
\multicolumn{3}{c}{Incoherent} & \multicolumn{3}{c}{Coherent}\\
\hline\hline
XRBs & TDEs & NSM & RRATs & FRBs & PSRs & GCRT & ESPs\\
stellar flares & SNe\tnote{a} & GRBs\tnote{a} & dMe flares & SNe\tnote{b} & GRBs\tnote{b} & EBHs & MSPs\\
% & & & EBHs & MSPs & GRBs\tnote{b} & SNe\tnote{b}\\
\hline
\end{tabular}
\begin{tablenotes}\footnotesize
\item [a] Afterglows
\item [b] Prompt
\end{tablenotes}
%\end{center}
\end{threeparttable}
\end{table}

\subsection{Fast Transients and Coherent Emission Processes}

Analogous to VLITE and LWA, both ngLOBO-High and -Low will have direct voltage sampling systems for detecting  fast ($<$1 second) transient phenomena down to msec timescales.

The unexplored region between VLITE and ngLOBO is much more exciting if violent, coherent emission processes are as common beyond our solar system as they are within it. Fig.~\ref{ngloboAREAL} includes purported coherent classes of emitters, including the Galactic center ``Burper'' (Hyman et al. 2005) detected by the same VLA feeds currently utilized by VLITE. The transient source marks at least one new class of coherently emitting sources that have been detected in the ngLOBO-High frequency range with current capabilities, but the pickings have been slim. 

Coherent source classes are listed in Table~\ref{tabTRANS}, and below we note two of the most important, magnetized extrasolar planets and millisecond pulsars (MSPs), also discussed in more detail in WGWP4. 

\subsubsection {Extrasolar Planets}

A prominent {\it predicted} class of coherent emitters are extra-solar planets, for example through the process of cyclotron maser emission. Their radio detection would be important in providing a new and direct method of exoplanet detection. Moreover, such signatures (present in all Solar system planets) would be a clear indicator of the presence of space weather-shielding {\it planetary magnetic fields}, a prerequisite to life as we know it. This is topic is discussed in more detail in Section 5.2, since the Jupiter analogue would favor the ngLOBO-Low frequency range. However, we note here that predictions for extrasolar planetary radio emission cover the entire ngLOBO frequency range (Lazio et al. 2004).

\begin{figure}[t!]
\begin{center}
\vspace{-1cm}
\includegraphics[width=5.5in,angle=0]{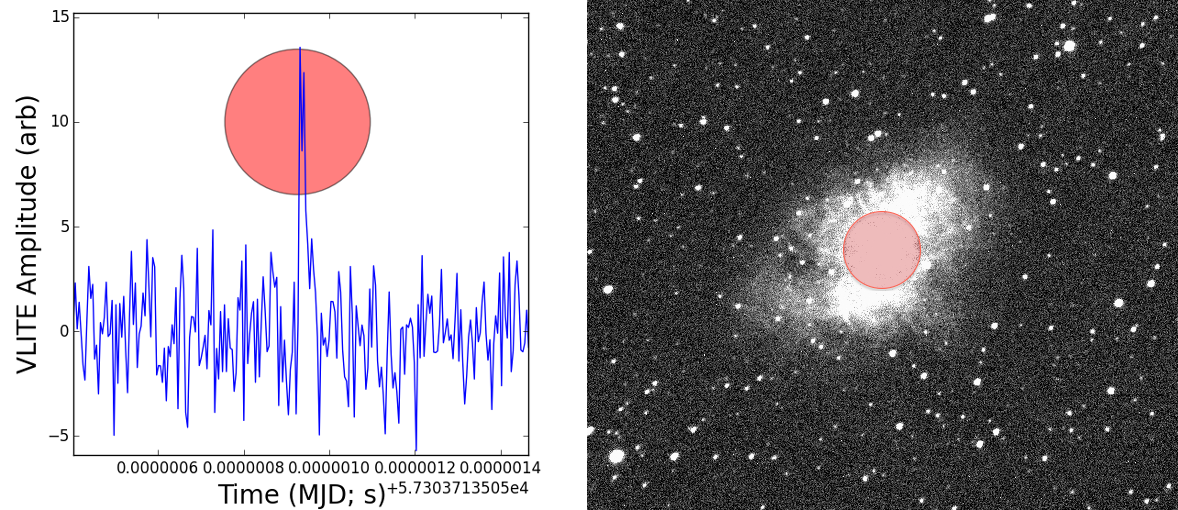}
\end{center}
\vspace{-0.5cm}
\caption{VLITE-FAST and {\it realfast} detection of Crab Giant Pulse. Left: Temporal association of {\it realfast} (L band, red circle) \& VLITE (350 MHz, blue) pulses. Right: Crab Giant Pulse association with realfast (L band, red circle) \& RAPTOR (optical, grey-scale). 
\label{realfast}}
\end{figure}

\subsubsection{Pulsars, RRATs, \& FRBs}

The current VLITE imaging correlator operates on second timescales. In parallel, a fast sampling system is being developed on VLITE, and should soon be useful for detecting PSR-like phenomena down to msec timescales. Fig.~\ref{realfast} shows a Crab Giant Pulse detected simultaneously in the time domain by VLITE at 340 MHz, and imaged at 1400 MHz by {\it realfast}.

%\todo{NEK: Kevin - can you say more about ngLOBO detecting PSRs and RRATs in its fast sampling mode? Keep it light and reference the suitable ngVLA Working Group White Paper?}

\begin{figure}[t!]
\begin{center}
\vspace{-1cm}
\includegraphics[width=5.5in,angle=0]{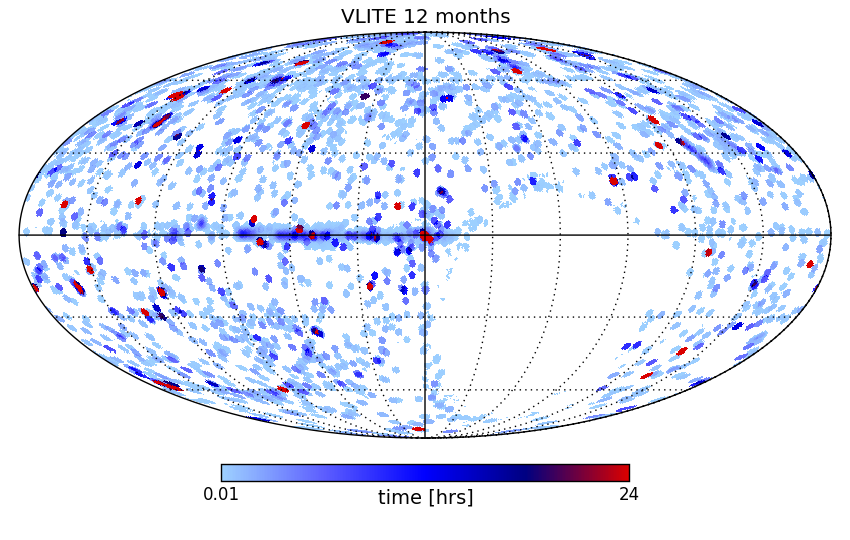}
\end{center}
\vspace{-0.5cm}
\caption{VLITE 12 month exposure map, with more than 6000 hours on sky.
\label{v12mon}}
\end{figure}

Projecting forward to ngLOBO, a fast sampling system should  be a formidable PSR and RRAT machine, for three key reasons. First, PSRs have extremely steep radio spectrum, even compared to extragalactic radio sources, as shown in Fig.~\ref{fig_msp}; a typical sweet spot is 400 MHz, striking a balance between brightness and both intrinsic and extrinsic turnovers. The second and third reasons are because of the coupling of ngLOBO's time on sky and FoV, respectively. At 400 MHz, the ngLOBO FoV will exceed 10 square degrees and, using VLITE as a guide, ngLOBO (Fig.~\ref{v12mon}) should yield $\geq$ 6000 hours on sky annually. In addition to excellent GHz and sub-GHz PSR science that has long emerged from instruments such as Parkes, the GBT, and the GMRT, several of the new aperture array instruments including the MWA (Murphy et al. 2017b, Bell et al. 2016) and LOFAR (Hessels et al. 2014, Stappers et al. 2011) continue to validate the contributions uniquely possible with thousands of hours of sub-GHz time on sky as available to ngLOBO annually. 

Fast Radio Bursts (FRBs) have now been detected down to 800 MHz. Rajmade and Lorimer (2017), considering the effects of thermal absorption and scattering, suggest FRBs might be detectable to 400 MHz or below. Moreover, low frequency detections offer unique constraints on FRB physical surroundings, through frequency dependent propagation effects. Moreover, since FRBs are likely cosmological (Chatterjee et al. 2017) these same propagation effects can serve as key diagnostics on the baryonic content of the Universe, much as PSRs probe the Galactic ISM today. 

An ngLOBO fast sampling system, and even VLITE, could detect bright FRBs if they are detectable below 500 MHz, even if band limited (Chowla et al. 2017). However, ngLOBO offers limited advantages over VLITE for current incoherent detection techniques. A more exciting approach is a fast imaging system for ngLOBO analagous to {\it realfast}. A GHz-band commensal imaging system on the ngVLA could play the same game - but it could not match ngLOBO's field of view. Moreover, the FRB $\log N-\log S$ distribution is observed to be shallow, indicating that a low gain/large FoV approach with ngLOBO is a good one. However, such an approach will demand computational power and coherent processing techniques beyond near-term capabilities, but very likely to exist in the ngVLA era.

%two key reasons: 1) Incoherent detection via dynamic spectra yields S/N $\propto N^{-1/2}$, where N is the number of antennas; and 2) while N = 300 for the ngLOBO, the dishes are smaller and their gain is lower than the VLITE/JVLA antennas. Conversely, ngLOBO has a tremendous advantage because of its large FoV and time on sky. However this only helps for detecting bright FRBs.

%direct on the timescale of the FRBs, as being pursued by {\it realfast} at L band with the JVLA, where the S/N $\propto N^{-1}$. VLITE is not capable of this approach because it lacks a fast correlator, and the problem is compounded by its limited instantaneous uv coverage. However, were ngLOBO equipped with a fast correlator it would be a good FRB detection machine, even at high DM. %, as shown in Fig.~\ref{frbNGLOBO}. 
%A GHz-band commensal imaging system on the ngVLA could play the same game - but it could not match ngLOBO's field of view. Coherent processing techniques, well beyond near-term capabilities, would empower ngLOBO fast transient detection significantly further in the ngVLA era.

By the ngVLA era, the initial excitement over FRB discovery may have subsided, but many more new questions will likely have been raised. More broadly, the discovery of FRBs may presage a much larger frontier of msec timescale, coherent processes rich for exploration, much as PSR phenomena have remained at the forefront of science decades after their initial discovery. 

\subsubsection{Millisecond PSRs and General Relativity}

Gravitational waves (GWs) are a critical prediction of General Relativity, and their detection by LIGO shook the world of physics. Matching in excitement are ongoing efforts to use a cosmic web of PSRs to detect lower frequency (nanohertz) GWs. Current detection efforts include the Nanohertz Observatory for Gravitational Waves (NANOGrav; Arzoumanian et al. 2015), the Parkes Pulsar Timing Array (Reardon et al. 2016), and the European Pulsar timing Array (Kramer \& Champion 2013).

\begin{wrapfigure}{R}{0.5\textwidth}[t]
%\begin{figure}[t]
\begin{center}
\vspace{-0.7cm}
\includegraphics[width=0.4\textwidth, angle=-90]{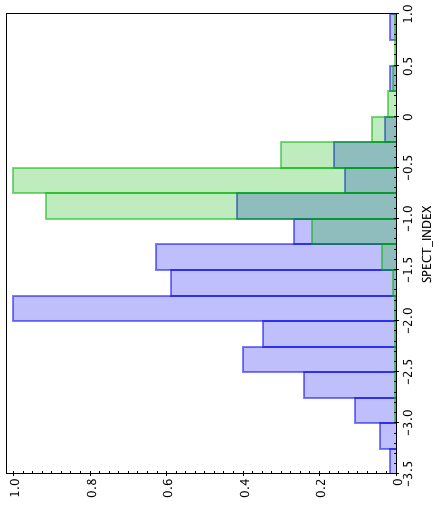}
\end{center}
\vspace{-0.6cm}
\caption{\small Normalized spectral index distributions for known pulsars (blue) and background radio sources (green). The median pulsar spectral index is $-1.8$ and that for radio sources is $-0.75$. The pulsar distribution is formed from 416 spectral indices taken from the ATNF PSRCAT, while the background radio source distribution is formed from approximately 1 million values from the TGSS and NVSS surveys.} \label{fig_msp}
%\vspace{-0.5cm}
%\end{figure}
\end{wrapfigure}

Experiments like NANOGrav rely on a pulsar timing array (PTA) comprised of steep-spectrum, millisecond pulsars (MSPs) that rotate with incredible stability, allowing them to be used as precise clocks. {\it Once identified}, these experiments require long-term monitoring of PTAs, with NANOGRAV now placing meaningful constraints on GWs from supermassive
black hole binary systems (Schutz \& Ma 2016). 

ngLOBO imaging will provide a powerful tool for increasing the census of MSPs. The technique dates to the discovery of the first MSP, that was presaged by analogy of its {\it steep radio spectrum} with the relatively fast spinning Crab pulsar (Backer et al. 1982, Mahoney and Erickson 1985); analagous detections in globular clusters followed (Erickson and Mahoney 1985b, Hamilton et al. 1985b). The approach works because steep-spectrum radio sources, {\it except for pulsars}, are rare. In Kimball and Ivezic (2008) fewer than 0.4\% of the radio sources have $\alpha< -1.5$ (where S$_\nu\propto\nu^\alpha$), while 2/3 of all known pulsars have such spectral indices (see Fig.~\ref{fig_msp}).  

Furthermore, since steep spectrum sources are naturally detected more efficiently at low frequencies, the FoV (e.g. $30'$ at 20-cm on the VLA) becomes an ally from the survey perspective. A beautiful demonstration is the recent detection of an MSP towards the GC (Bhakta et al. 2017) from the TGSS ADR1 survey at 150 MHz (Intema et al. 2017). This {\it hybrid MSP search} approach is indirect, but after {\it localization} though low frequency imaging, steep-spectrum sources can efficiently be followed up with a single-dish telescope at frequencies high enough to avoid propagation affects (Bhakta et al. 2017). Alternatively, accurate imaging positions can be used to search for gamma-ray pulsations (Clark et al. 2017). Hybrid imaging has now been used to successfully identify six MSPs toward  \emph{Fermi} sources, including the second fastest known (e.g. Frail et al. 2017).  

The detection of GWs from MSPS also requires the
removal of the effects of the time variable interstellar medium (ISM) through measurements of the dispersion measure (DM). ngLOBO will provide the high DM precision accessible at low frequencies, as demonstrated by LWA and LOFAR (Stovall et al. 2015).  ngLOBO-high, in particular, hits a sweet spot for detecting MSPs; a regime in which they are bright but above where intrinsic turnovers often occur.  

%\subsubsection{PSRs and Related Phenomena} 

%$\bullet$ PSRs as probes of the ISM: 
%Independent of GR applications, ngLOBO will probe ISM electron density and magnetic fields through measurements of time variable dispersion, scattering, and Faraday rotation in the large set of normal pulsars that will be accessible.  

%$\bullet$ {Rotating Radio Transients (RRATs)}: RRATs are an example of exotic neutron star systems accessible to both ngLOBO-Low and ngLOBO-High.

\vspace{-0.5cm}
\section{Commensal Science: ngLOBO Imaging of ngVLA Targets}
\vspace{-0.25cm}

%\tododetail{Namir: Needs work. Pick out a few exciting examples of how commensal observations help ngVLA targeted science. Few others in Appendix?}

Distinct from its independent operation as a Radio LSST, ngLOBO will provide broad-band, low-frequency images of all sources targeted by ngVLA, creating a database of information for many of the most interesting sources in the Universe. The lower frequencies will be sensitive to steep-spectrum non-thermal emission and structures that are resolved out at the higher frequencies, while the broad frequency coverage will offer detailed spectral information and the opportunity to search for HI absorption and molecular emission and absorption lines. The comparatively large low frequency field of view will vastly extend the database to cover all sources in a $>$ 1\deg ~radius (increasing with decreasing frequency) along each site-line observed. This additional data can be mined for serendipitous discoveries as well as targeted Galactic and extra-Galactic science.

%Distinct from its independent operation as a Radio LSST, ngLOBO can often compliment ngVLA science through complementary low frequency images. Some will leverage ngLOBO's ability to provide a better constraint on the distribution and spectrum of nonthermal emission from larger ngVLA targets, for example in supernova remnants, normal galaxies, radio galaxies, and clusters. ngLOBO will also be sensitive to thermal emission and absorption. The former, when combined with ngVLA images, is useful for separating the distributions of relativistic particles and ionized gas in a wide variety of sources, through their continuum spectra. The case of absorption can be even more interesting, by providing distance information through delineating the relative radial superposition of thermal and nonthermal constituents along common lines of sight, or by observing HII regions as discrete absorption regions against the Galactic background.  

Below we review a few areas of complimentary ngLOBO Galactic and extragalactic commensal science. These examples are limited to continuum observations. A discussion of ngLOBO spectroscopic science, much of which is exploratory in nature, is reviewed in the Appendix. 

% Here we present a compendium of non-time variable continuum and spectral line science, informed mainly from the ngVLA Science Working Group White Papers, where complimentary sub-GHz images could be useful. A few select topics are expanded upon in the Appendix. Where pertinent, a reference is made to the appropriate ngVLA White Paper whence the topic is discussed in the context of ngVLA.

\subsection{Commensal Galactic Science}

\begin{figure}[ht!]
\begin{center}
\includegraphics[width=5.5in,angle=0]{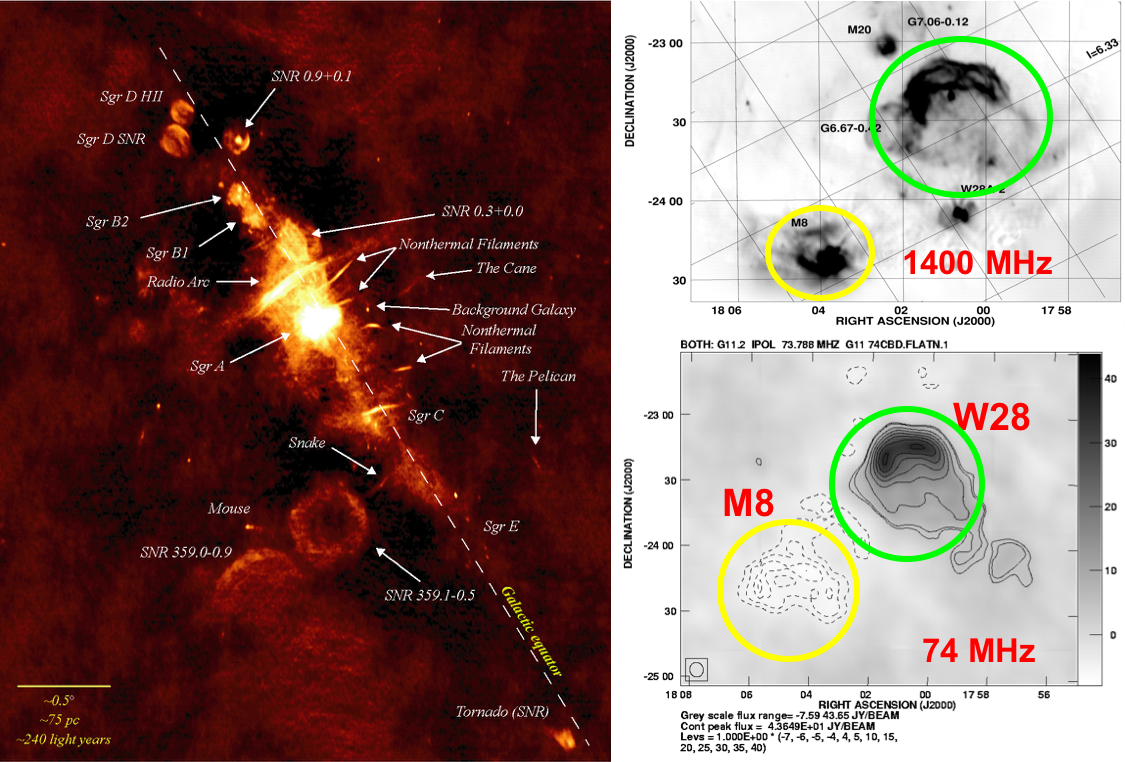}
\end{center}
\vspace{-0.5cm}
  \caption{Left: 330 MHz VLA image of the Galactic center (LaRosa et al. 2000). Both HII regions and SNRs are in emission at this intermediate ngLOBO frequency. Right: The upper panel is a 1400 MHz VLA image of the SNR W28 (green circle) and the HII Region M8 (yellow circle), both in emission. At lower ngLOBO frequencies, the HII region goes into absorption, as shown in the lower panel (Brogan et al. 2003).
}
\label{GC}
\end{figure}

Sub-GHz Galactic observations across the ngLOBO frequency range will be a powerful compliment to many higher frequency Galactic targets. The 330 MHz wide-field Galactic center image (Fig.~\ref{GC}, left, LaRosa et al. 2000) illustrates many of these, as follows:

\subsubsection{SNRs and their interaction with the ISM}

The ngLOBO frequency range is ideal for discovering SNRs, such G0.3+0.0 (Frail and Kassim 1999, discovered in Fig.~\ref{GC}), while Brogan et al.\ (2006) detected more than 30 new SNRs further up the plane, also at 330 MHz. Sensitive images at ngLOBO frequencies can achieve a complete census of Galactic SNRs, essential to understanding Galactic star formation history and ISM energy input. SNR generated cosmic rays represent $\sim$1/3 of the interstellar medium (ISM) energy density, drive Galactic chemistry through the ionization of atoms and molecules. 

\subsubsection{Cosmic Ray Acceleration}

Nearly all models for SNR cosmic ray acceleration by shocks or turbulent Fermi processes predict subtle radio spectral index variations. When complimented by higher frequency ngVLA observations, spectra can be studied to high accuracy because of the long leverage arm in frequency space. Applications pertain to the measurement of thermal absorption, that probe the intervening (Kassim et al. 1989), surrounding (Brogan et al. 2005), and unshocked ejecta {\it within} young SNRs (Delaney et al. 2015). Moreover, accurate, spatially resolved spectra can quantify curvature and other subtle predictions of shock acceleration theory (Reynolds and Ellison 1992, Anderson and Rudnick 1996). Wide-field, low frequency Galactic observations are naturally serendipitous and can uncover rare sources tying cosmic ray acceleration from radio to $\gamma$-rays (Brogan et al. 2005)

\subsubsection{Thermal Emission and Non-thermal Exotica}

Aside from classic SNRs and PSRs, the high surface brightness sensitivity to non-thermal emission reveals sources or source components not revealed at higher frequencies; many non-thermal filaments, including the Pelican (Lang et al. 1999), were discovered at 330 MHz (Fig.~\ref{GC} left) before being followed up in earnest at ngVLA frequencies. ngVLA will observe many Galactic HII regions. For ngLOBO, many of these (e.g. the Sgr D SNR, Fig.~\ref{GC} left) will be optically thick in emission. When combined with ngVLA, this can offer physical constraints for electron densities, temperatures, and filling factors (Kassim et al. 1989). 

\subsubsection{Discrete Thermal Absorption}

Fig.~\ref{GC} (right) also illustrates how ngLOBO-Low aperture array measurements offer unique information for Galactic cosmic-ray astrophysics. The top panel shows an inner Galaxy HII region and SNR, both in emission at ngVLA frequencies. The bottom panel shows both sources at the ngLOBO-Low frequency of 74 MHz, where the SNR remains in emission while the HII region goes into absorption against the Galactic background (Brogan et al. 2003). The measurement of discrete absorption provides a powerful, 3-D constraint on the distribution of the cosmic ray electron gas in the Galaxy (Nord et al. 2006). When combined with gamma-ray observations, e.g. from Fermi and future gamma-ray observations from space, the degeneracy between electron density and magnetic field from radio measurements alone can be lifted (Longair, 1990).  A comprehensive summary could quantify SNRs as the primary source of cosmic rays in the Milky Way (Longair 1990, Webber 1990). Impressive MWA observations of discrete HII absorption are emerging (Hurley-Walker et al. 2017), and measurements to lower frequencies are crucial for inferring the energy spectrum of the emitting particles. LOFAR has limited access to the inner Galaxy where many of the most important lines of sight lie, and SKA Low will likely not extend sufficiently low in frequency.

\subsection{Commensal Extragalactic Science: Particle Acceleration and Ageing}

\begin{figure}[ht!]
\begin{center}
\includegraphics[width=6.75in,angle=0]{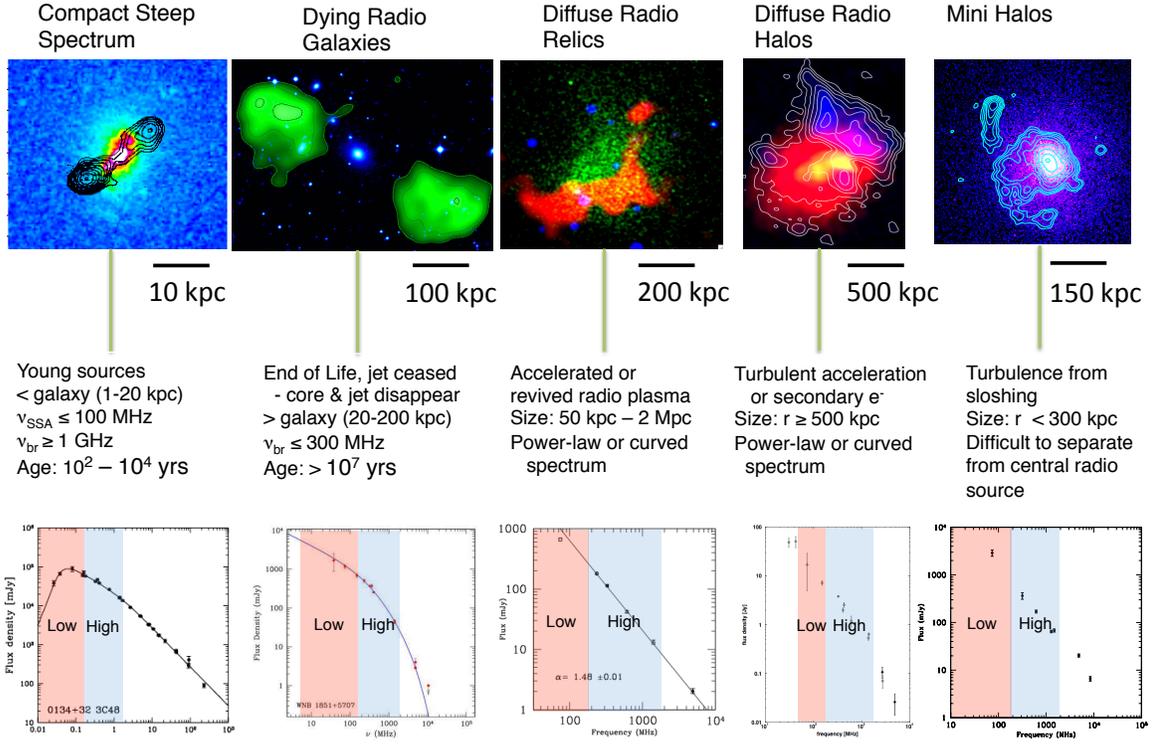}
\end{center}
\vspace{-0.95cm}
  \caption{Examples of extragalactic source populations that are well suited to low frequency observations. The spectral signatures of the sources, including breaks and turnovers,  reveal details of the particle acceleration and aging mechanisms.
}
\label{ngloboSCIENCE}
\end{figure}

The combination of ngLOBO with ngVLA will provide simultaneous frequency coverage 
from 20 MHz to 2 GHz (with ngLOBO) up to 100 GHz (with ngVLA) at arcsecond (or better) 
resolution and excellent sensitivity. This will allow us to study with unprecedented 
detail and spectral coverage a number of extragalactic non-thermal phenomena. 
Fig.~\ref{ngloboSCIENCE} provides examples of commensal extragalactic science, 
that includes studies of the lifecycle of radio galaxies from their early active 
stage ($10^2-10^4$ yrs) to highly evolved fading/dying phases ($>10^{7-8}$ yrs), and 
studies of the origin and evolution of diffuse radio sources in galaxy clusters 
(radio relics, halos and minihalos).

\subsubsection{Radio galaxies and environment}

Understanding the origin and evolution of the synchrotron-emitting relativistic electrons in 
radio-loud active galaxies has great interest in extragalactic astronomy. Long-standing 
questions in radio-galaxy physics are how the nuclear activity in the central AGN is triggered, 
whether the environment plays a role, and how the AGN energy input, transported by jets and 
stored in the lobes, is eventually transferred to the external medium in the host galaxy and 
cluster environments (Sect.\ 4.2.3). 

Low-frequency observations play an important role in the study of the radio galaxy phenomenon. 
Given their roughly power-law radio spectra with a steep index ($\alpha < -0.5$), low-frequency 
observations are particularly sensitive to the extended, evolved components (lobes) of radio galaxies, 
whose particle energy density is dominated by low-energy electrons. Low-frequency observations 
are also sensitive to the injected spectrum of the relativistic particles (Begelman, Blandford \& Rees 1984, 
Rieger et al.\ 2007). As a matter of fact, low-frequency observations are essential to identify 
turnovers in the spectra of those regions where the acceleration of relativistic electrons is 
believed to occurr, i.e., the compact hot spots at the jet end in high radio luminosity sources and the jets in low power radio galaxies. Finally, low-frequency observations can trace the final stages of the life of a radio galaxy, when the mass accretion onto the supermassive black hole of the host galaxy is interrupted or becomes insufficient to feed the nuclear activity, and the radio source enters a dying phase (e.g., Murgia et al.\ 2011). At these final stages, all components produced by continuing activity (core, jets and hot-spots) will disappear and the lobe emission will rapidly fade, accompanied by a dramatic steepening of its radio spectrum ($\alpha \sim -2$; 
Fig.~\ref{ngloboSCIENCE}). Low-frequency observations are therefore well-suited 
to search for these faint, ultra-steep spectrum radio sources. However, the number of dying radio galaxies known to date is limited to a handful, likely due to the poor sensitivity of the current low frequency surveys. 

The combination of sensitivity and high angular resolution offered by ngLOBO with allow 
us to separate out the cores, jets, lobes and hot spots of radio galaxies. The simultaneous 
higher frequency observations at a matched resolution of 1-10 arcseconds will allow us to 
obtain a well-constrained integrated radio spectra and detailed images of the spectral index distribution, thus probing the physical conditions of the sources across a large interval
of electron energy and in different evolutionary stages. This will include studies of 
the population of still elusive fossil radio sources, which ngLOBO is expected to 
unveil. The detection of low-frequency turnovers in jets and hot spots will 
provide information on the still very uncertain shape of the electron spectrum at low 
energies ($\gamma \sim 100$) and thus constrain particle acceleration mechanisms.

One of the main statistical tools to investigate the effect of the environment on the
AGN activity is the radio luminosity function for galaxies in different environments. 
Despite the wealth of X-ray, optical and IR data, this research has been limited 
mainly by the statistical information in available radio survey data. ngLOBO 
will enable the exploration of the faint end of the radio luminosity function 
and separation of the AGN and starburst contribution, allowing the study of the
environment on the population of low and high luminosity radio galaxies.

\subsubsection{Particle Acceleration in Dynamically Complex Clusters: Radio Halos and Relics}

A striking feature of a number of galaxy clusters is the presence of diffuse non-thermal 
radio emission of very low surface brightness ($\mu$Jy/beam) and size that may reach and 
exceed a Mpc. This extended, steep-spectrum ($\alpha < -1$) emission takes the form of 
large peripheral radio relics and centrally located radio halos (an example of a cluster 
hosting both a central radio halo and a relic is shown in Fig.~\ref{ngloboSCIENCE}; see 
Feretti et al.\ 2012 for a review). The importance of halos and relics is nowadays widely 
recognized because their existence requires $\mu$Gauss magnetic fields and ultrarelativistic 
electrons to be distributed throughout a large fraction of the cluster volume (Brunetti \& 
Jones 2014 for a review).  Radio halos and relics are typically associated with dynamically 
disturbed clusters with evidence for recent or ongoing cluster merger activity (e.g., Cassano et al.\ 2010, 
Yuan et al.\ 2015). Based on this association, the origin of these sources has been linked 
to the (re)acceleration of cosmic ray electrons in the intracluster medium (ICM) and 
amplification of the cluster magnetic field by large-scale shock waves and turbulence 
driven by cluster-cluster collisions. A number of correlations relating non-thermal to 
thermal cluster properties (i.e., radio luminosity to cluster X--ray luminosity or mass) 
have been found for clusters with diffuse radio emission (e.g., Cassano et al.\ 2013, de 
Gasperin et al.\ 2014), further supporting to the connection between halos/relics and 
cluster mergers.

The fraction of clusters hosting a radio halo and/or a relic is still uncertain. Current 
statistical analyses on the occurrence of radio halos are limited to clusters with 
very high X-ray luminosity and mass ($M_{500}>6 \times 10^{14}$ $M_{\odot}$) at low/intermediate 
redshifts ($z<0.4$). Roughly $60-80\%$ of the most massive of these systems (those with mass
$> 8\times 10^{14}$ $M_{\odot}$) are found to host a radio halo, while this fraction appears 
to decrease to $\sim 20-30\%$ at lower masses (Cuciti et al. 2015). The fraction of clusters 
with a radio relic is estimated to be about $5-10\%$ in high X-ray luminosity clusters 
(e.g., Kale et al. 2015) and, like halos, there is indication that this fraction may 
increase with the cluster X-ray luminosity (van Weeren et al. 2011).

ngLOBO will work to address the remaining gaps in our understanding of diffuse 
radio emission in dynamically disturbed clusters, through deep commensal observations of 
these complex merging environments and statistics from a large commensal survey of thousands 
of clusters over a wide range of cluster masses and redshifts. Its excellent sensitivity 
will allow us to probe the presence of halos and relics in lower mass clusters, to detect 
the first diffuse radio sources that may form at high redshift and to constrain the 
number density of halos/relics and its evolution with cosmic time. ngLOBO will also be
be uniquely sensitive to ultra-steep spectrum ($\alpha < -1.5$) radio halos, 
sources with a high-frequency cutoff in their synchrotron spectrum at $< 1$ GHz and thus detected only at
lower frequencies (e.g., Brunetti et al. 2008). This class of radio halos is expected to
originate via turbulent reacceleration of electrons during less energetic cluster collisions 
(e.g., between two low-mass clusters or a massive cluster and a much less massive subcluster), 
which are thought to be ubiquitous in the Universe. ngLOBO has the potential to discover ultra-steep 
spectrum halos in hundreds of galaxy clusters with no diffuse emission currently detected 
at GHz frequencies. ngLOBO will therefore probe the role of shocks and turbulence in 
the formation and evolution of large-scale structures in the Universe and place tight constraints on the 
cosmological evolution of the largest Dark Matter halos in the Universe. ngLOBO will 
also provide a unique opportunity to probe the strength and structure of extragalactic magnetic 
fields on Mpc scales.

X-ray observations of clusters probe the Dark Energy density and equation of state by 
measuring the baryonic mass fraction of the Universe, but they depend on the identification 
of a large, relaxed cluster sample, a costly and often ambiguous task at optical and 
X-ray wavelengths. ngLOBO-High observations will be an efficient method for distinguishing 
a relaxed sample of galaxy clusters for Dark Energy studies.

%ngLOBO-High will be uniquely sensitive to the diffuse radio emission associated with merger-driven 
%turbulence and shock waves that heat the intercluster medium (ICM) and compress its magnetic fields. 
%By tracing the steep-spectrum emission of radio halos and relics from thousands of clusters over a 
%wide range of redshifts, ngLOBO-High will accurately trace cluster number density and merger frequency, 
%placing tight constraints on the cosmological evolution of the largest Dark Matter halos in the Universe. 
%ngLOBO-High images will also be able to differentiate magnetic-field configurations and particle-acceleration 
%mechanisms in clusters.

\subsubsection{Galaxy Cluster Cores: AGN Feedback and Radio Minihalos}

Early expectations for the centers of galaxy clusters were that they should contain 
massive amounts of cooling gas which was vigorously forming stars (Fabian 1994). 
High resolution spectroscopic X-ray observations of clusters, however, did not reveal 
the predicted catastrophic cooling at the cluster centers (e.g., Peterson \& Fabian 2006). 
This led to the idea that clusters must contain a source of heating which is distributed 
through the cluster core and roughly balances the cooling from thermal Bremsstrahlung 
emission of the hot gas. The most promising candidate for the gas heating is the 
supermassive black hole at the heart of the central brightest cluster galaxy (BCG). 
Statistical studies have shown that the BCG in cool-core clusters (clusters where 
the core cooling time is shorter than the Hubble time) is active in the radio in nearly 
100\% of strong cool cores and $\sim 70$ \% of weak cool cores, whereas the fraction 
of radio-emitting BGCs drops to $\sim 45$ \% in non-cool core clusters (Mittal et al. 2009). 
These numbers suggest that the central AGNs are responding to the massive amounts of 
gas cooling in the cluster cores through increased activity in the radio band, leading to 
an energy feeback process in which heating and cooling are tightly coupled. 
AGN feedback may also play a crucial role in shaping the luminosity function of the 
host galaxy (Croton et al. 2006) and contribute to pre-heating of the ICM (Dubois et al. 2011). 

High-resolution images of the bright X-ray emission from the cluster cores have revealed 
the presence of X-ray cavities, which contain little to no thermal emission, but are 
often filled by synchrotron radio emission associated with the central BCG 
visible at GHz frequencies and below. These radio-filled cavities are evidence that the radio jets and 
lobes driven by the central supermassive black holes are pushing aside the thermal gas 
and depositing energy into the ICM to offset radiative losses and to suppress the 
catastrophic cooling predicted by the cooling flow model (e.g., Fabian 2012  
for a review). A primary uncertainty in this feedback process is how the AGN jet
energy is transferred and distributed into the thermal gas and the degree to which 
other heating mechanisms (thermal conduction, sound waves, turbulence) are contributing 
(e.g., Fabian et al. 2006, Zhuravleva et al. 2014).

%The mechanical power of AGNs which have launced jets of radio emission 
%into the ICM may also be responsible for driving turbulence into the cluster. The turbulence will 
%dissipate into heat to offset the radiative cooling. As seen in the well-studied Perseus cluster 
%({\bf REF}), the AGN appears to also drive sound waves into the ICM to heat the gas. Relativistic 
%particles and magnetic fields from the central AGN are distributed through the cluster core 
%during the outbursts.
%One additional source of ICM turbulence in the cores of cool core clusters is sloshing of the cool gas 
%in the central region of the cluster's gravitational potential. This sloshing can be driven by tidal 
%disruptions from passing dark matter subhalos. The turbulence introduced by the sloshing motions can 
%re-accelerate the (sub-)relativistic particles that the central AGN has mixed within the thermal gas. 
%Evidence for this re-acceleration comes from the observations of radio mini-halos in dynamically 
%disturbed ool cores ({\bf Simona REF}). 

In addition to AGN-inflated X-ray cavities, X-ray observations have also revealed sloshing motions 
of the gas in the cool cores of most relaxed clusters, resulting from a gravitational perturbation 
of the cluster central potential in response to collisions with small subclusters (Ascasibar \& 
Markevitch 2006). These motions, observed as sharp spiral- or arc-like cold fronts (Markevitch \& Vikhlinin 2007), 
may amplify the local magnetic field and generate turbulence in the cool core (e.g., ZuHone et al. 2013). 
Turbulence, in turn, may contribute to balance the radiative cooling in the cluster core and, 
at the same time, re-accelerate aged relativistic electrons injected by past activity of the central 
radio galaxy and mixed with the thermal gas. Evidence for this re-acceleration may come from the 
observations of radio minihalos --- diffuse, low surface brightness radio sources with steep radio 
spectrum ($\alpha < -1$) --- in the sloshing cool cores of relaxed clusters (e.g., Giacintucci et al. 2014;
an example of minihalo is shown in Fig.~\ref{ngloboSCIENCE}). 

Observations at frequencies around 1 GHz of cool-core clusters often reveal both the central extended active 
radio source and the surrounding diffuse radio minihalo, if present. However, it is generally difficult 
to disentangle the minihalo emission from the radio galaxy, as this requires high sensitivity radio data 
over a range of resolutions and frequencies, and thus studies of large samples of clusters with minihalos have only 
recently been possible (e.g., Giacintucci et al.\ 2017). It became clear that, contrary to general belief, 
minihalos are a fairly common phenomenon in the cool core of massive clusters, whereas there is indication 
for a drop of their fraction in cool-core clusters with lower total masses. However, large amounts of
uncertainty still exist concerning the origin and physical properties of minihalos, their relation to 
the cluster core dynamics and turbulence, their connection to the activity (past and/or current) of 
the central radio galaxy and to the gas heating processes that are quenching cooling flows.

ngLOBO will have the required sensitivity, resolution, and frequency coverage to study a 
large population of radio galaxies in a broad range of radio luminosity, thus probing the feedback 
mechanisms whereby radio galaxies impact the formation and evolution of groups and clusters 
of galaxies, and potentially suppress star formation in large elliptical galaxies, 
leading to the observed exponential cut-off in the bright end of the luminosity function.
ngLOBO low-frequency observations will be essential to trace the extended (10-100 kpc) diffuse 
and faint aged radio lobes, best detectable at frequencies of few hundred MHz, with an 
active radio nucleus, detected at higher frequencies. At the same time, it will allow 
us to identify older relativistic particle populations from past outbursts of the central 
AGN, confined in ``ghost'' cavities (e.g., the two outer X-ray cavities by Fabian et al.\ 2002). 
Despite the many uncertainties, the study of the radio spectrum in the aged and 
active components can provide reliable information on the cycles of activity of the 
BCG and constrain the total AGN energy output injected into the ICM over the cluster 
lifetime (e.g., Clarke at al.\ 2009, Giacintucci et al.\ 2012).

The combination of sensitivity and high angular resolution offered by ngLOBO will also 
allow us to explore the minihalo phenomenon in a much larger sample of galaxy clusters of 
various masses and cosmological distances. We will probe the formation of minihalos in 
clusters with lower mass and constrain the minihalo occurrence as a function of cluster 
mass and redshift. Complementary to future X-ray measurements, this will provide a
unique opportunity to probe turbulent motions in the very central regions of 
cluster atmospheres. A recent direct measurement of the gas velocities in the cool core
of the Perseus cluster, which hosts a radio minihalo, with the Hitomi X-ray satellite 
revealed the presence of mild turbulence sufficient for reaccelerating relativistic 
electrons and possibly for balancing the radiative cooling in the core 
(Hitomi Collaboration et al. 2016).

\subsubsection{Nature of the Earliest Active Galaxies}

Emission from the most powerful and distant high redshift-radio galaxies (HzRGs) is a critical signpost of the collapse of protogalaxies in the early Universe. HzRGs are among the most luminous, massive, and distant objects in the Universe, usually identified by steep-spectrum radio emission powered by super-massive black hole (SMBH) accretion. They are energetic sources across the whole electro-magnetic spectrum revealing diverse components in the proto-cluster environment and providing important diagnostics of key physical constituents in the early Universe, including relativistic plasma, hot and warm ionized gas, cool atomic gas, molecular gas, dust, old and young stellar populations, quasars and SMBHs. The most efficient way to identify possible HzRGs is to search for ultra-steep spectrum objects in low-frequency sky surveys. The extension of frequency coverage of the ngVLA below 1 GHz is crucial to identifying these objects at high redshifts, particularly if they exist beyond $z \geq 8$. As signposts of Dark Matter potential wells, ngLOBO-High identifications invite natural optical/IR and radio line (e.g. CO) follow-ups to search for groups of galaxies.

\subsection{ngLOBO and the EOR}

The excruciating effort to detect a global EOR signature is a prerequisite to opening the Epoch of Reionization (EOR) for follow-on radio imaging. It is increasingly clear this effort is best pursued by dedicated experiment-class instruments either looking for the spectral signature with single dipoles such as EDGES (Bowman et al. 2008, Monsalve et al. 2017), or looking for the signature in the power spectrum with interferometers such as HERA. Here we focus on HERA. While there is frequency overlap between HERA and ngLOBO-High, there is no baseline overlap because the spatial scales are vastly different. Moreover, pursuing both the cosmological and high angular resolution imaging goals of low frequency astronomy through a single instrument has proven to be problematic. 

Taking a longer term vision suggests a path towards a future convergence. From a cosmological perspective, the emission HERA targets corresponds to global-scale structure in the very early Universe (z~$>$~6). Over Hubble time-scales they evolve into smaller scale structures detected by ngLOBO-High, e.g. radio galaxies and clusters. These data could be indirectly beneficial to HERA, as they represent a contaminating foreground signal HERA, and follow-on EOR imaging experiments, hope to avoid. A similar argument could be made for ngLOBO-Low and Dark Ages experiments such as LEDA and DARE.

%Between the the largest and smallest scales accessible to ngLOBO (clusters) and the EOR experiments (e.g. Zeldovich pancakes), respectively, lies a cosmological gulf of Large Scale Structure (LSS). Realizing an instrument to fill this gap requires an "ngLOBO-High Core" or alternatively "ngHERA". The ngVLA infrastructure provides an excellent location to embed this capability. At that point, HERA, ngLOBO, and a low frequency ngVLA/ngLOBO core would provide a full comological panorama from the EOR to the present. A similar argument exists for a condensed ngLOBO-Low Core to bridge the gap to the global-scale structure detected by the Dark Ages experiments, e.g. LEDA and DARE, at even higher redshifts.  

\vspace{-0.5cm}
\section{Independent Science with Aperture Arrays}
\vspace{-0.25cm}

At these lowest frequencies observable from the ground
we are making exciting discoveries
about the radio afterglows from fireballs entering the
atmosphere (Obenberger et al.\ 2014, 2015) pulsars (Stovall et
al.\ 2015), and many other topics within and beyond our solar system
(Taylor et al.\ 2012 and references therein); as of July 2017 the Long
Wavelength Array (LWA) has produced 39 refereed
publications. There
are also now other passive uses such as studies of the ionosphere, plasmasphere, lightning, solar and Jovian bursts, and space weather. Below we
describe two key science areas.

\vspace{-0.5cm}
\subsection{Fireballs}
\vspace{-0.25cm}

\begin{figure}[ht!]
\begin{center}
%\vspace{9cm}
%\special{psfile=fireballv2.eps hoffset=-15 voffset=-40 hscale=50.0 vscale=50.0 angle=0}
%\special{psfile=FB1.eps hoffset=270 voffset=-20 hscale=35.0 vscale=35.0 angle=0}
\includegraphics[width=5.5in,angle=0]{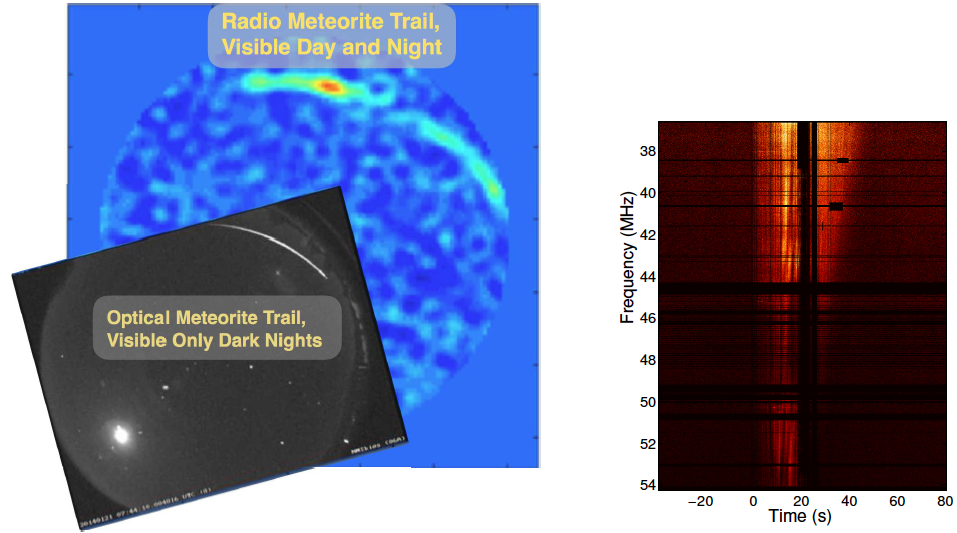}
\end{center}
\vspace{-0.5cm}
  \caption{
{\bf Left:} All-Sky Image from the LWA1 at 38 MHz (color), and the optical image from the NASA All-Sky Fireball Network (inset bottom-left), showing the same bright elongated fireball viewed from a location in Southern New Mexico. {\bf Right:} Dynamic spectrum of the self-emission from a typical fireball.
}
\label{fireball}
\end{figure}

A surprise from LWA1 was the discovery of strong (kiloJy)
low frequency self-emission from large meteors (Obenberger et al.\ 2014; 
2015; 2016; 
Fig.~\ref{fireball}).
The light-curves of these events
typically show a fast rise followed by a slower decay, lasting up to
several minutes, with steep spectral properties (Obenberger et al.\ 2015). 
Our current hypothesis is
that the emission is due to the radiation of Langmuir waves within
the trail; the frequency of the emission would therefore be
proportional to the square root of the electron density, providing a
useful probe of the trail plasma. Thus far, 160 events have been
discovered in $\sim$20,000 hours of all-sky image data below 50 MHz, and
none in 3,000 hours above 50 MHz. Besides opening up new ways to 
study meteors, an advantage over optical detectors is that, with the LWA, 
we can observe day or night, rain or shine.  

These radio techniques will be used to complement Lidar measurements
of the upper atmosphere that have allowed for the study of winds,
turbulence and the vertical transport of elements (Gardner \& Liu
2010).  These transport mechanisms are of interest since they
influence the layered structure of the Earth's atmosphere.  Our radio
observations will be sensitive to very small motions ($\sim$1 m/s),
but only when there is a bright meteor available to probe the
atmosphere.  As the geometry permits we can also take advantage of
specular reflections from transmitters on the ground to measure the
Doppler shift and transverse motions of the ionization trails left by
meteors.  Helmboldt et al. (2014) used the LWA1 in this way and found
velocity measurements of $\sim$100 m/s for the winds.

\vspace{-0.5cm}
\subsection{Extra-Solar Planets and Moons}
\vspace{-0.25cm}

Jupiter is well known to emit powerful bursts of coherent emission
that can out-shine all other sources below 40 MHz.  In fact all of the
planets in our solar-system with magnetic fields have been observed to
produce coherent emission, with the maximum frequency linked to the
field strength.  This raises the intriguing possibility of not only
detecting extra-solar planets by way of their low frequency emission,
but also using the emission characteristics to measure the magnetic
field strength.  Magnetic fields are of critical importance to the
retention of planetary atmospheres and thus the development of life.
Recent evidence from the MAVEN spacecraft suggests that Mars lost much
of its atmosphere due to its lack of a significant magnetic field
(Jakosky et al. 2015).  From the modulation of the radio bursts it
would also be possible to directly obtain the rotation period of the
planet.  By looking for other periodic signatures we might also 
learn about the presence of moons.  Some of the
strongest radio bursts from Jupiter are associated with Io.

While many searches have been carried out (Winglee et al.\ 1986, Lazio
et al.\ 2004, Murphy et al.\ 2015 to name just a few), no radio bursts
from exo-planets have yet been detected.  However, almost all of these
searches have been executed at frequencies well above Jupiter's cutoff
at 40 MHz.  Searches with the existing LWA1 facility have been plagued
by high confusion noise causing a lack of sensitivity.  One way around this is increasing the baseline lengths as has been done at OVRO-LWA (Hallinan, priv. communication).  We also plan to address
this by combining the two New Mexico LWA stations, thereby doubling the sensitivity
and effectively applying a 
spatial filter to reduce the confusion noise by several orders of
magnitude.  We estimate our 10 minute
sensitivity at 30 MHz for two LWA stations at 50 mJy which should 
allow us to see a Jupiter-like planet out to $\sim$3 pc (but farther if
the exoplanet is more luminous) assuming that we are
inside the beaming cone. This is high-risk, high-reward science that
relates to the successful Kepler mission and was one of the most
prominent areas of research endorsed by the most recent decadal 
review - ``New Worlds, New Horizons in Astronomy and Astrophysics''.

\vspace{-0.5cm}
\subsection{Space Weather, Solar, and Sun-Earth Connection}
\vspace{-0.25cm}

The importance of scientific understanding of Coronal Mass Ejections (CMEs), especially their roles in the development of the solar corona and their space weather effects at the Earth, is well known and well documented (see, for example, the introductory book on CMEs by Howard (2011)). It is widely recognized that magnetism plays a critical role in the formation, launch, and early evolution of CMEs, and that the orientations of CME fields govern the strength of magnetic storms at the Earth. Understanding the nature of the core magnetic field of CMEs, especially when they are close to the Sun, is therefore central to our knowledge of their formation, their contribution to the development of the solar environment, and their impact on space weather.

Measurements of this intrinsic magnetic field have been highly elusive over the decades, because no in-situ spacecraft has yet passed through the solar corona, and because traditional methods for remotely imaging the corona do not provide a means by which the field can be measured. We have recently demonstrated that low frequency observations of pulsars passing behind the sun can be used to probe the density and solar wind of CMEs (Howard et al. 2016).  Faraday rotation will be the only reliable and regularly available means by which this measurement can be made. Despite a small number of isolated studies, this has never been performed on a consistent and systematic scale. The ultimate objective is to combine the latest instrumentation, latest data processing pipelines, cutting edge analytical tools, and a cross-disciplinary task force, to make systematic measurements of the intrinsic density and magnetic field of near-Sun CMEs, and track their behavior during their passage through the corona.

Regarding the Heliophysics Decadal Survey, this work applies directly to Key Science Goals 1 (origins) and 3 (interaction). Further, the Diversify (D) and Realize (R) components of the DRIVE initiative both include recommendations for support for ground-based systems. This effort would also contribute to Challenge SHP-3 (magnetic energy) and note that multidisciplinary cooperative efforts are listed as a Top Level Priority (Priority 2.0) of the Decadal Survey. Finally, this project directly addresses Objectives H1 (solar plasmas and magnetic fields), H2 (variability), W1 (boundary conditions), W2 (predict activity), and W3 (predict disturbances) of the 2014 NASA Heliophysics Roadmap.

Looking beyond our solar system, with the sensitivity improvements of ngLOBO it may be possible to study the space weather in exoplanet systems.  Space weather can have a profound impact on the climate and habitability of exoplanets.  Low frequency observations measuring the energetics of flares and CMEs in stars hosting planetary systems could complement the work being done at optical and UV wavelengths.

\subsection{Ionosphere and Plasmasphere Applications}

The ionosphere can strongly affect the radio emission from cosmic sources.  For this reason alone it is worth understanding so that its effects may be removed.  If the ngVLA extends to L band, ngLOBO ionospheric characterization will enhance polarization calibration and dynamic scheduling. Beyond that, there is a large scientific community interested in understanding the ionosphere and plasmasphere based on intrinsic scientific merits. There is also a large portfolio of scientific applications including but not limited to navigation, communication, space weather prediction, and space situational awareness, derivable from precision ionospheric remote sensing afforded a low frequency interferometer like ngLOBO.  A few of these are discussed below. Needless to say many of these applications bring with them the possibility of attracting not only a new community, but also independent resources to a combined ngLOBO/ngVLA facility.

\subsubsection{Ionosphere} The ionosphere is a thin shell of cold plasma that exists along side neutral particles in the upper atmosphere of the Earth. It is created by interaction of radiation, mostly from the sun, and the neutral particles of the air (Budden 1985). The majority of the ionosphere is created from extreme UV radiation, notably the Lyman series of lines from the elements in the Sun, which are absorbed at a range of heights. The radiation is most intense at the highest altitudes where the air is thinnest, and ionizes the air almost completely. As the radiation is attenuated, its flux diminishes and the density of air increases, further increasing the absorption rate. As such, the majority of deposition of a type and frequency of radiation happens in a single band in the ionosphere, which generates the characteristic banding of the ionosphere. The plasma remains in these bands since they are far outnumbered by the neutral atmosphere at these altitudes, and are very collisional. At higher altitudes where the ionosphere is thinner, the particles co-rotate with the Earth, trapped in flux tubes associated with its magnetic field. Beyond the plasmasphere, electric fields driven by the solar wind dominate over co-rotation, forming the dynamic region known as the magnetosphere.  Since the creation of the ionosphere is generally due to the sun, the ionosphere has a strong seasonal, diurnal and solar cycle dependence. A small fraction of the ionosphere is created from cosmic rays impacting the atmosphere, so it is never absent, but it fluctuates greatly with solar activity.  Of considerable interest is  understanding the  detailed structure of the ionosphere and improving the global models that are used to represent it.

\subsubsection{Earth, Ionosphere, and Space Weather Science}

As a specific example, low frequency interferometry provides the most accurate measure of turbulence and small-scale ionospheric waves (wavelengths $\lesssim$50 km), through precisions measurements of differences in total electron content ($\Delta$TEC). VLITE $\Delta$TEC measurements are 2-3 orders of magnitude more accurate than GPS based measurements.  First pioneered at the VLA at 330 MHz (Jacobson and Erickson 1992)  with the discovery of a new class of traveling ionospheric  disturbances (TIDs), researchers are still only scratching the surface of Earth science applications that would be possible with ngLOBO. 

Expanding upon initial results from the early 1990s, researchers found that the superb sensitivity of the VLA to plasma density fluctuations made it capable of picking up signatures of density gradients between flux tubes in the plasmasphere.  In fact, the new class of magnetic eastward directed ``TIDs'' discovered by Jacobson and Erickson (1992) were later determined to be such plasmaspheric irregularities that appear as field aligned wavefronts due to co-rotation.  With the advent of the 74 MHz system on the VLA (Kassim et al. 2007), the additional capability of monitoring several sources at once was added given the field of view is more than four times larger at 74 MHz than at 330 MHz.  This allowed for high-precision specification of ionospheric fluctuations over a larger area.  Using data from the VLA Low-frequency Sky Survey (VLSS; Cohen et al. 2009), Helmboldt et al.\ (2012) showed a variety of wavelike disturbances present near the VLA with wavelengths between $\sim$50--100 km.  While some shared characteristics with well-known classes of TIDs, the nature of many is still unknown.

\begin{figure}
\includegraphics[width=\textwidth]{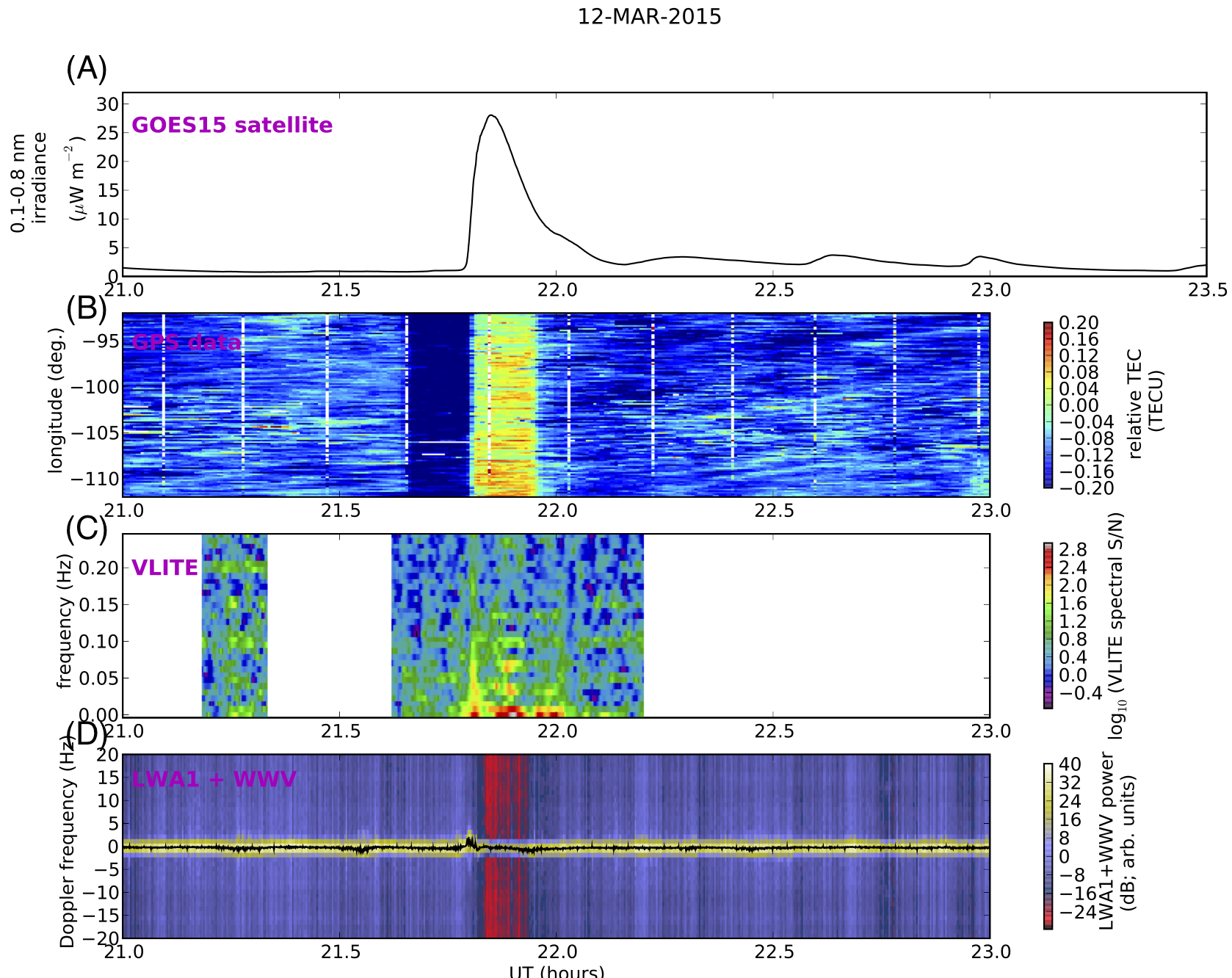}
\caption{From Helmboldt et al.\ (2015), data characterizing the impact of an M-class solar flare from 12 March 2015: (A) X-ray irradiance from the GOES15 satellite; (B) Mean, de-trended TEC from GPS stations in the continental United States as a function of longitude and time; (C) TEC gradient fluctuation amplitude from VLITE as a function of oscillation frequency and time; (D) 10-MHz Doppler spectrum of the ionospheric reflection of the WWV carrier signal at a receive site near the VLA.  Note the VLITE-detected fluctuations are consistent with plasmaspheric irregularities about 1,300--4,000 km from the VLA.}
\label{flare}
\end{figure}

Simultaneous measurements toward a single bright source and several moderately bright sources with the same 74 MHz field of view by Helmboldt and Intema (2012) were used to characterize both ionospheric and plasmaspheric disturbances along the VLA line(s) of sight.  This type of analysis enables a unique capability to study connections between plasmaspheric and ionospheric activity, especially when combined with other sources of ionospheric remote sensing data.  An example of this was shown by Helmboldt et al.\ (2015) when the impact of an M-class solar flare was observed with the VLA/VLITE, an oblique HF ionospheric Doppler sounding experiment, and GPS receivers.  While the HF sounder and GPS data showed clear signs of ionospheric disturbances on local and hemispheric scales, respectively, VLITE observations toward a bright source (3C84) showed the signature of brief plasmaspheric irregularities as well (see Fig.\ \ref{flare}).

\subsubsection{Remote Sensing Applications}

There are a growing number of applications being pursued by the ionospheric and plasmaspheric remote sensing community, many of them being pioneered by the LWA and VLITE. As first beautifully characterized by the MWA (Loi et al.\ 2015), VLITE now regularly affords a real-time, tomography-like radial characterization of plasmaspheric flux tubes. In turn, such measurements can be utilized for applications related to Over-The-Horizon (OTH) radar and geolocation. Another project underway with LWA and VLITE is the demonstration of remote detection of underground explosions, in conjunction with the Energetic Materials Research and Testing Center (EMRTC) at NM Tech. Other applications for ngLOBO include solar radar for accurate geomagnetic storm prediction, passive radar for many other applications, and for satellite tracking, terrain mapping, space debris tracking, and detection of Near Earth Objects.

\vspace{-0.5cm}
\section{Capabilities of the Instrument}
\vspace{-0.25cm}

\subsection{ngLOBO-Low Capabilities and Specifications}

The frequency range for ngLOBO-Low would be 5-150 MHz.  The lower 
end is set by the interests of ionospheric studies and given the 
strength of the signals involved antenna requirements are less stringent than for radio astronomy.  Astrophysical studies would be carried out between 20-150 MHz
where it is anticipated that the performance will be sky noise
dominated.

The ngLOBO-Low system we are proposing to co-locate $\sim$50 LWA-style stations with ngVLA antennas to provide arc second resolution and mJy sensitivity for 5 to 150 MHz.   Since LWA stations are currently optimized for observations between 20 and 80 MHz, some design effort is required to extend the frequency coverage up to 150 MHz.  In particular, both the antenna, front end electronics, and analog receiver designs will need to be updated. A gap of 20 MHz between 88 and 108 MHz would
be unusable due to FM broadcast; alternatively this band could be tapped for passive radar applications and Earth science studies.

Additional development is needed on the signal processing side as well
to increase the frequency coverage up to 150 MHz.  This work is
facilitated by the recent development of the Bifrost framework which
has been implementated on commodity hardware at LWA-SV (Cranmer et
al. 2017).  For ngLOBO-Low we would want to either increase the number
of beams (currently 4 on LWA1 each with two tunings) or the bandwidth
of the beams (currently 20 MHz for each tuning). The goal would be to
enable full coverage of the spectrum in at least two directions
simultaneously as is currently the case for the LWA over a smaller
frequency range.  This dual beam setup would also allow the LWA
stations to support PI-driven projects in addition to providing
commensal capabilities to ngVLA projects.  The end result would
provide $\sim$mJy sensitivity at arcsecond level resolution (see
specifications in Table~\ref{tabSPECS}).

\begin{table}
\begin{center}
\caption{ngLOBO-Low Specifications}\label{tabSPECS}
\vspace{0.2cm}
\begin{tabular}{lc}
\hline
\hline
Field-of-view for beams at 80 MHz (arcmin) & 120\\
Number of Beams & 8\\
Number of tunings/beam & 2\\
Effective Area (m$^2$) & 65\,000 \\
GND Frequency range & 20-150 MHz \\
Beam Bandwidth (MHz) & 20 \\
SEFD (single beam) & 6 kJy\\
Beam rms in 1 hour (mJy/beam) & 0.64 \\
Resolution FWHM (arcsec) & 3 \\
\hline
\end{tabular}
\end{center}
\end{table}

\subsection{ngLOBO-High Capabilities and Specifications}

ngLOBO-High will access the $\sim$ 150 - 800 MHz radiation field
accessible near the prime focus of the ngVLA antennas. The lower
frequency end will dovetail into the upper frequency role off of
ngLOBO-Low. ngLOBO-High will extend as high as possible in frequency,
approaching 1 GHz at reduced performance if technically feasible.  
Because the system is uncooled the relative sensitivity at
the higher end will be modest but still viable for brighter
source classes, e.g. Lorimer-class FRBs or bright ngVLA (or field)
targets.

\subsection{Technical Concept Options}

There are various options that need to be explored for the receiving system (receivers + feeds) capable of meeting these technical performance goals. 

Receivers: System temperatures as low as 30K are now routinely
achievable for uncooled receiver systems.  A commensal ngLOBO-High
receiver system could be front-end noise dominated to above 500 MHz
(H. Mani, private communication). The sky temperature ranges from 350
K at 150 MHz to 0.5 K at 2 GHz. Receiver design does not pose a
significant technical challenge or risk.

Feeds: This is the most significant area of technical development required, and is not addressed by the VLITE, LWA, or LOBO precursors. Detailed studies must await the final ngVLA design convergence, since feed performance depends critically on its design. In order to be commensal, all options would need to sample the radiation field near the prime focus of ngVLA dishes. Since these are likely to be an offset Gregorian design, the simplest concept would incorporate omni directional dipole feeds using the subreflector as a backplane. Alternatively, unidirectional log periodic feeds could be advantageous in not requiring a backplane. A more sophisticated design could involve using the subreflector structure itself as a feed.  Once the ngVLA dish design is firm, we will conduct electromagnetic simulations (e.g. in CST, as used by NRL for the LWA antenna design) to characterize candidate feed concepts.

In Table~\ref{tabSPECShi} we provide Working estimates of the specifications for ngLOBO-High scaled to 300 MHz from ``Next generation VLA Notional parameters'' at:
\url{https://science.nrao.edu/futures/ngvla/concepts} which are themselves still in flux. 

\begin{table}
\begin{center}
\caption{ngLOBO-High Specifications (300 MHz)}\label{tabSPECShi}
\vspace{0.2cm}
\begin{tabular}{lc}
\hline
\hline
Field-of-view (arcmin) & 193\\
Aperture Efficiency (\%) & 30\\
Effective Area (m$^2$) & 31000 \\
System Temperature (K) & 50 \\
Bandwidth (GHz) & 0.5\\
Continuum rms in 1 hour ($\mu$ Jy/beam) & 7.15\\
Resolution FWHM (arcsec) & 0.93\\
\hline
\end{tabular}
\end{center}
\end{table}

\vspace{-0.5cm}
\section{Synergies and Lessons Learned}
\vspace{-0.25cm}

\subsection{Dipole Arrays}

Below frequencies of $\sim$150 MHz dipole arrays are more
cost-effective than dishes, to the point where infrastructure costs
(buildings, power and communications) begin to make up a significant
fraction of the total cost. Dipole arrays have additional advantages in
that they have much larger fields-of-view which can be exploited for
surveys, ionosphere and transients, though at the expense of
complexity in electronics and software.

The LWA1 station was co-located with the VLA in part to reduce
infrastructure costs (very short power and fiber runs), and in
part with the idea of eventually combining the two instruments.
The LWA concept involves 50 stations spread around New Mexico
and it has long been considered that once established, these 
stations could also host dishes that expand the capabilities of
the VLA.  Once baselines of 300 km are under consideration, the
high cost of fiber and power distribution could become a significant
fraction of the expense for dipole arrays or dishes.  Procuring 
land can also be expensive as it generally requires archaeological
and environmental studies.  

\subsubsection{Fiber}

Throughout the development of the LWA project, we have learned much
about procuring land and running power and fiber in rural New Mexico
where RFI conditions can be quite good. 

There are a number of modest sized telecommunications companies (e.g. Western New Mexico Communications headquartered in Silver City, NM) that have hundreds of miles of high-volume fiber optic cable.  In most cases direct access to the fiber is not permitted, and one is limited to leasing a connection.  By way of example a 1 Gbps connection between the VLA and Socorro costs roughly \$50K/year.  Given the extremely high data rates expected for the ngVLA, and the location of antennas well removed from access points, it is clear that NRAO will need to operate its own fiber network.  To partially defray the large costs involved it may be possible to set up partnerships with school, hospitals, police and fire departments, and other public agencies that have a need for high speed communications.  Substantial Federal subsidies are available to schools.  New Mexico currently boasts some of the highest costs/bit for its more rural schools which has  led
to the current New Mexico Broadband Initiative to increase access to broadband.  Currently New Mexico ranks 42nd in the nation on the list of most connected states\footnote{broadbandnow.com}.

We have
learned that while running fiber over power lines is much cheaper than
burying it (a decision made for the $\sim$40 km fiber run from Socorro
to Sevilleta at a cost of only $\sim$\$1.50/foot compared to typical
buried fiber costs of \$5/foot), in some rural areas
the lead weights placed on the cables are used for target practice by
the locals with the result that the fiber is sometimes broken.
Combined with outages due to high winds and thunderstorms it may be
more cost effective, and much more reliable, to bury the fiber despite
the larger initial investment. As discussed above, a substantial savings can be realized by sharing the  constructions costs with other public agencies. 

\subsubsection{Power and Land}

The cost of running power is even greater than fiber on a per foot basis, although access to power is generally available throughout New Mexico.  For a 1 mile power run the cost charged by the utility is roughly \$100K.   Power and fiber are not allowed to share the same conduit though there can still be a cost savings in the installation of both in the same trench.  

\begin{figure}[ht!]
\begin{center}
\vspace{-1cm}
\includegraphics[width=5.75in,angle=0]{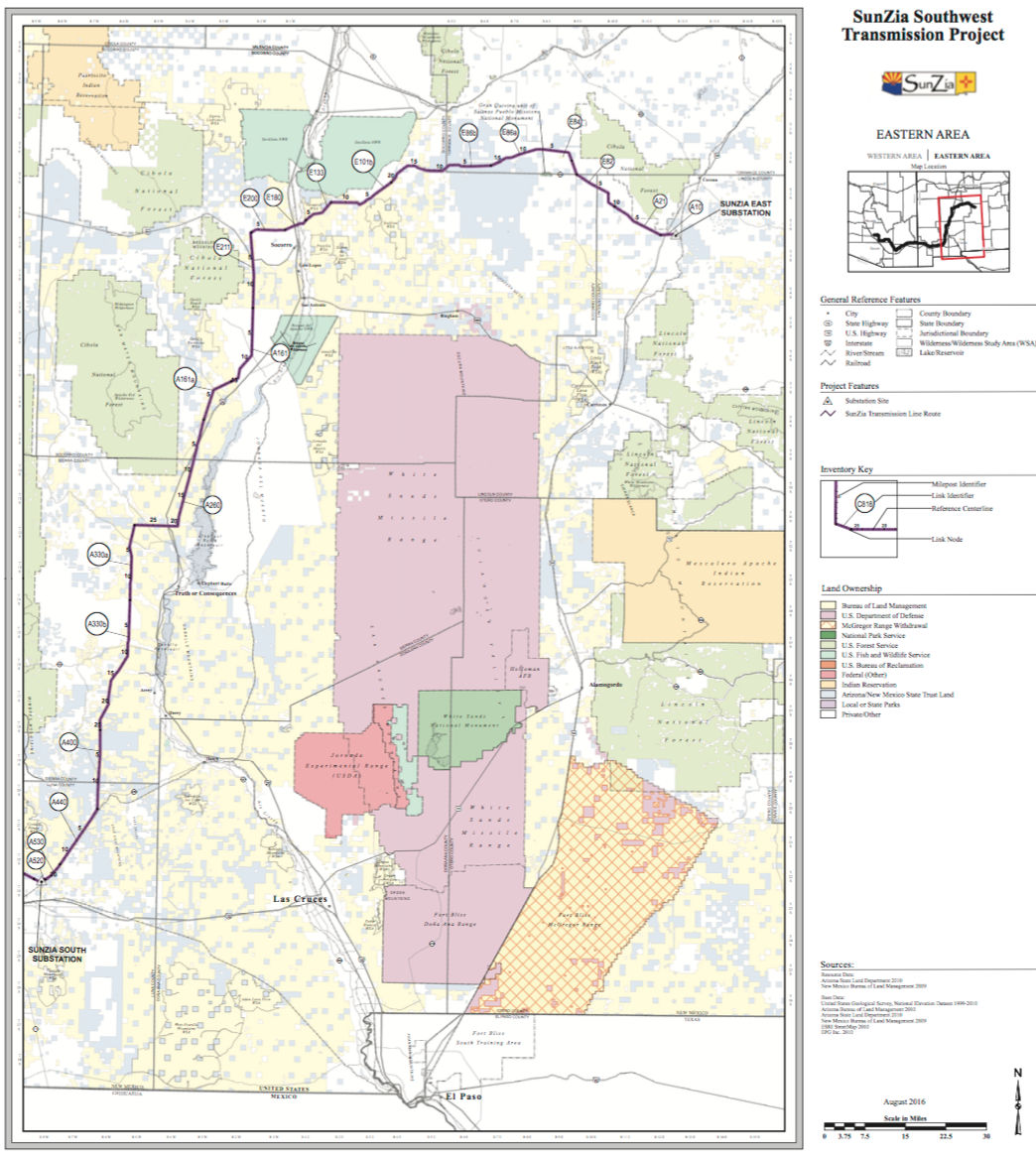}
\end{center}
\vspace{-0.5cm}
  \caption{
Map of the proposed route for the SunZia Southwest Transmission Project.
}
\label{sunzia}
\end{figure}

Overhead power lines are a source of unwanted Radio Frequency Interference (RFI).  For this reason LWA stations have a requirement of burying at least the last 0.5 miles to the station.  The degree of RFI scales with the voltage of the power line such that high voltage lines are to be avoided.  A study of expected RFI from high voltage power lines showed that the distance between the antenna and a high-voltage line should be at least 10 miles (Crane 2010).  This creates a problem in that there are a number of companies contemplating how to deliver power from windfarms in eastern New Mexico to power markets in Arizona and California.  One example is
the SunZia Southwest Transmission Project which is planning to build two 
300 kv AC lines (see Fig.~\ref{sunzia}.  Configurations for the ngVLA will want to avoid this corridor.

Outside of federal wilderness areas and BLM land, much of the land in
rural New Mexico is under the control of the State Land Office. We
have had many discussions with the New Mexico State Land Office and
understand some of the options for establishing new leases, as well as
the benefits of having a University partner.  The Universities in New Mexico are major beneficiaries of the revenue generated by the State Land Office. 

\subsubsection{Automation}

In order to improve efficiency and reduce operating costs for the LWA we
have developed a high degree of automation at the stations.  This is
accomplished at LWA1 using the Heuristic Automation for LWA1 (HAL)
system.  The principal duties of HAL are:  the creation and
implementation of daily observing schedules, monitoring of the station
health, and responding to target of opportunity observations.

For scheduling, HAL creates a schedule that contains a combination of
targeted observations from PIs, lower priority ``filler'' observations,
and passive station monitoring checks.  The schedule is created several
hours before the start of the UT day in order to allow review by LWA
staff.  After observations have completed HAL also handles the movement
of the observation metadata to the LWA Archive.  Automatic  data
transfers to the LWA Archive and the LWA Users Computing Facility are also
scheduled as part of the metadata management process.  These copies are
implemented in an ``operationally aware'' way such that they can be
interrupted and resumed as necessary.

HAL also checks a variety of monitoring points in order to actively
protect the station.  These include both external (environmental)
factors, e.g., nearby lightning, and internal (equipment) factors, e.g.,
overheating due to HVAC failure.  In the event that a situation could
adversely affect the station, HAL will automatically shutdown critical
components.  For external shutdown events, HAL will also restore the
station to an operational state once the event has cleared.  For
example, the station will be restarted after a thunderstorm clears the
area and no lightning has been detected for at least 30 minutes.

Finally, HAL implements the automatic trigger processing necessary for
target of opportunity observations.  When a trigger is received, the
processing is implemented such that existing observations are gracefully
interrupted and new observations submitted.  The HAL system also
automatically notifies any PI that their observations have been
interrupted or cancelled due to a trigger.  By moving to this system the
LWA has seen a reduction in the trigger response time, from notification
to being on sky, from one to two hours with a manual system to one to
two minutes.  The data from the triggered observations can 
also be automatically transferred as with standard observations.

An automated system such as this would be beneficial for not only
ngLOBO but ngVLA as a whole.  This is particularity true given the large
number of antennas and monitoring points per antenna.  Furthermore,
external factors, such as weather, will be different over the wide
geographic area that the ngVLA will cover and the system as a whole must
be able to cope with antennas dropping out and returning to the array.
Finally, automation on each antenna will also minimize the need for
manual intervention and reduce the possibility of potential damage in
situations where external control of the antenna is not possible.  These
features can lead to a reduction in the staffing and maintenance costs
of the array.

\subsection{Commensal Observing}

Commensal observing programs that are tied to following observations designed by the PI of the higher frequency observing bands do not necessarily obtain independent calibration sources. In the case of VLITE, we relied on having a sufficiently bright calibrator observed with sufficient frequency at each primary observing band to undertake the calibration. Commensal observing can also place a large burden on archival storage and computer processing where the 24/7 operation can lead to very large data volumes in parallel to the primary observing program data. Finally commensal observations can lead to challenges when it comes to data rights, particularly for the area of the sky immediately overlapping with the primary observer field of view.

\subsubsection{Commensal Calibration}

The calibration of a fully slaved system such as ngLOBO-high naturally depends on the calibration undertaken by the higher frequency observing programs. For the VLITE system, we explored the NRAO VLA archive observing programs for one year, splitting data by primary observing band and investigating calibration trends. This showed that for any observing band over a 24 hour period there was nearly always a sufficiently long observation of a calibrator to permit accurate calibration of the low frequency data. In practice for VLITE we have found there to be very few programs where we are unable to obtain a calibrator within our processing window. We have also been exploring bandpass stability of the VLA and found that antennas that undergo no maintenance are stable for months at a time. This opens the avenue to development and application of global bandpass models to the commenal data where it may be possible to largely eliminate the need for calibrators if the system is sufficiently stable.

Looking forward to ngLOBO, calibration of the low frequency system should be taken into account as the ngVLA operational concept is developed. If the ngVLA system is designed to be very stable over long periods and rely on regular calibration observations taken by the observatory then it will be important to be sure that there are appropriate low frequency calibrator sources included within that program. Also important to consider under calibration are observations of low frequency polarized sources of known polarization angle (e.g. pulsars) to allow full polarization calibration of ngLOBO.

\subsubsection{Processing Commensal Data}

Challenges to commensal data processing are significantly reduced if the number of options in the imaging procedure are reduced or eliminated altogether. This choice is relatively less restricting at longer wavelengths because wide fields contain hundreds of similar radio sources and in most cases can be considered generically. A single VLITE pipeline routinely produces high quality images, with the exception of rare, pathological fields, e.g. containing the Sun, Galactic center, or one of the bright A-team sources. 

It is also extremely beneficial for automated pipelines to output both calibrated ($u,v$) data and images. For VLITE, on occasions when a generic image does not suffice, the calibrated ($u,v$) data can usually be re-imaged manually but at greatly reduced overhead since all the basic calibration and flagging steps have already been applied.

\subsubsection{Commensal Data Proprietary Period}

Based on our experience with VLITE, we suggest two generic domains: 1) If the PI explicitly expresses no interest in the low frequency commensal data, it should be made immediately available to all users; 2) If the PI exercises proprietary data rights, it should remain so on a timescale similar to that for all higher frequency ngVLA data. 

Obviously much more complex arrangements will almost certainly be called for. For example, a PI may observe a radio galaxy with the ngVLA, and request proprietary data rights for the portion of the ngLOBO commensal image centered on their target. However another user, whose interest is in either quiescent or transient sources within the larger commensal field of view unrelated to the PI's target should not be barred from access to the data. The details of how these, and a myriad of other complex cases, are handled should be taken into account as a subset of data rights considerations for the observatory as a whole.

Even more nonstandard cases relate to applications of the data falling outside radio astronomy. For example, VLITE runs an ionospheric reduction pipeline in parallel to its radio astronomy imaging one, and the two have almost nothing in common from the data rights perspective. It would seem unreasonable and certainly inefficient for ngLOBO data deemed proprietary by the PI for radio astronomy purposes, to be made unavailable to applications in ionospheric science. Many other cases, both imagined and as yet undefined, for example accessing the data stream from a direct voltage sampling system for satellite tracking or passive radar applications, should be considered almost completely independent of the radio astronomy data.

\vspace{-0.5cm}
\section{Configuration}
\vspace{-0.25cm}

\begin{figure}[t!]
\begin{center}
\vspace{-1cm}
\includegraphics[width=3.0in,angle=0]{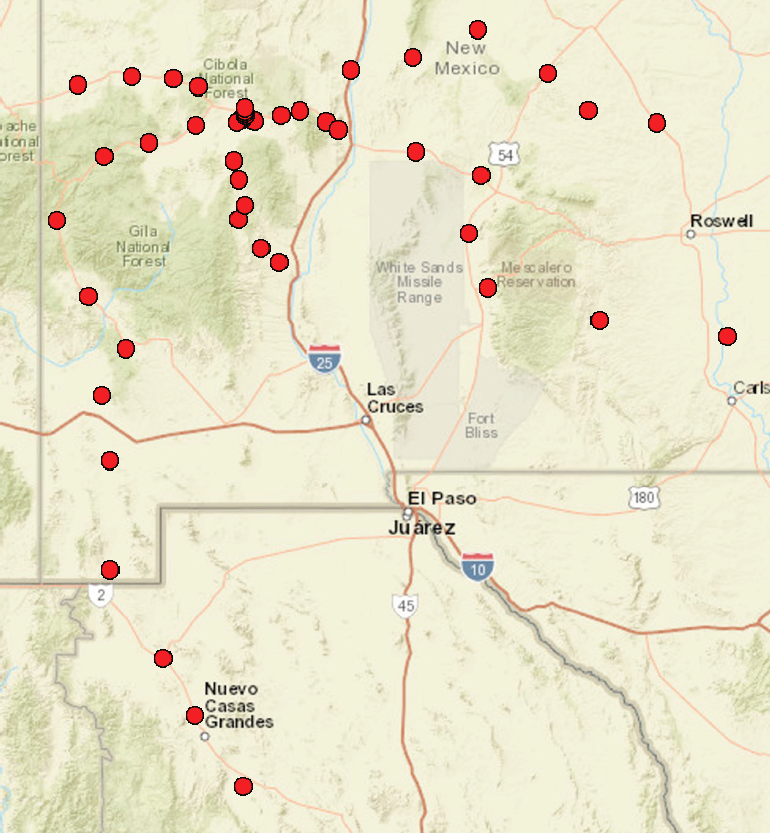}
\includegraphics[width=3.0in,angle=0]{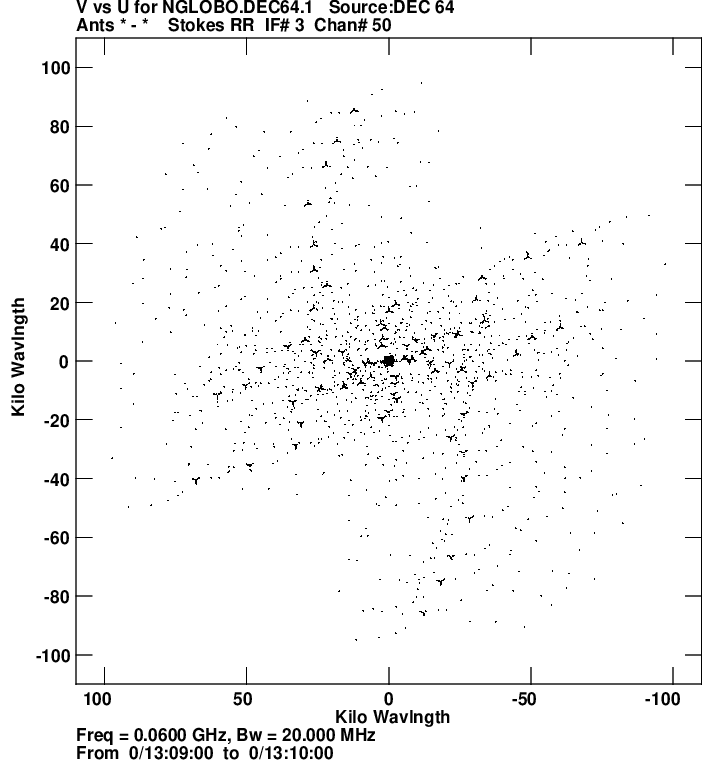}
\includegraphics[width=3.0in,angle=0]{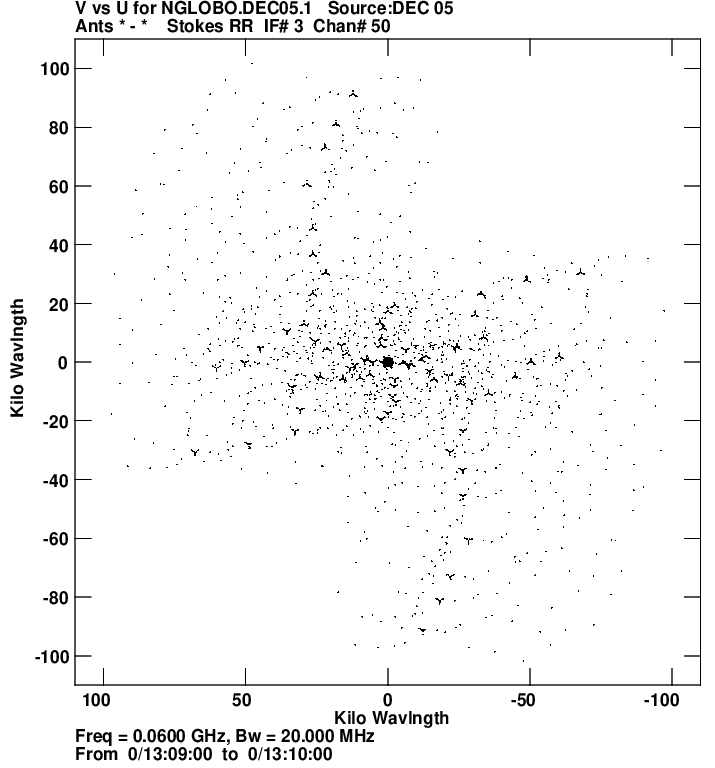}
\includegraphics[width=3.0in,angle=0]{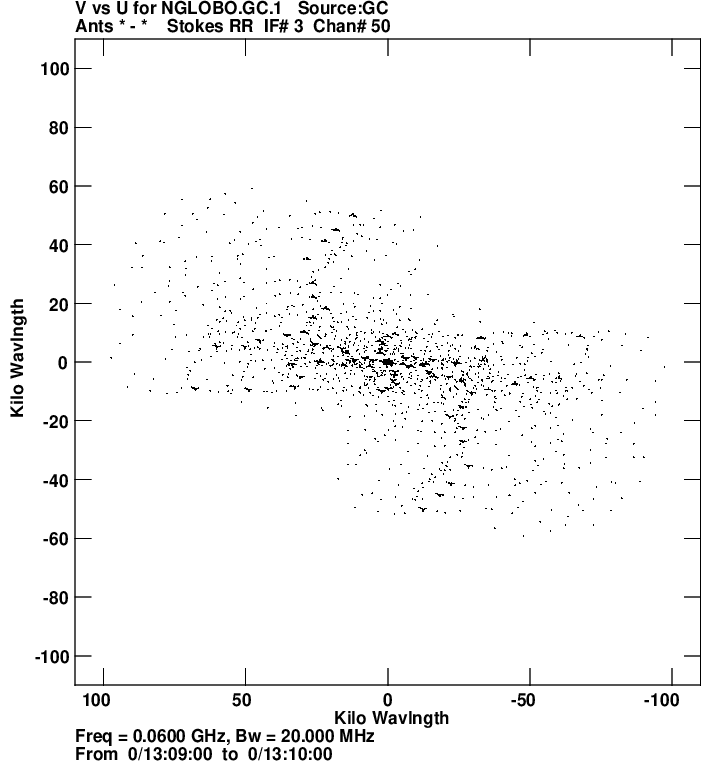}

\end{center}
\vspace{-0.5cm}
  \caption{
ngLOBO-Low configuration and the resulting snapshot  ($u,v$) coverage for a source at declination +64 (top-right), +05 (bottom-left),
and $-$29 (bottom-right).  The observing frequency is 60 MHz but bandwidth synthesis is not enabled.
}
\label{snapshot}
\end{figure}

Assuming 50 stations for ngLOBO-Low we obtain the snapshot ($u,v$) coverage shown in Fig.~\ref{snapshot} for sources at declinations of +64, +05, and $-$29 (corresponding to the declination of the Galactic Center). This configuration employs 13 VLA pad positions from A configuration and 37 locations that have been tentatively identified as possible ngVLA sites including LWA-SV.  

\begin{figure}[t!]
\begin{center}
\vspace{-1cm}
\includegraphics[width=3.0in,angle=0]{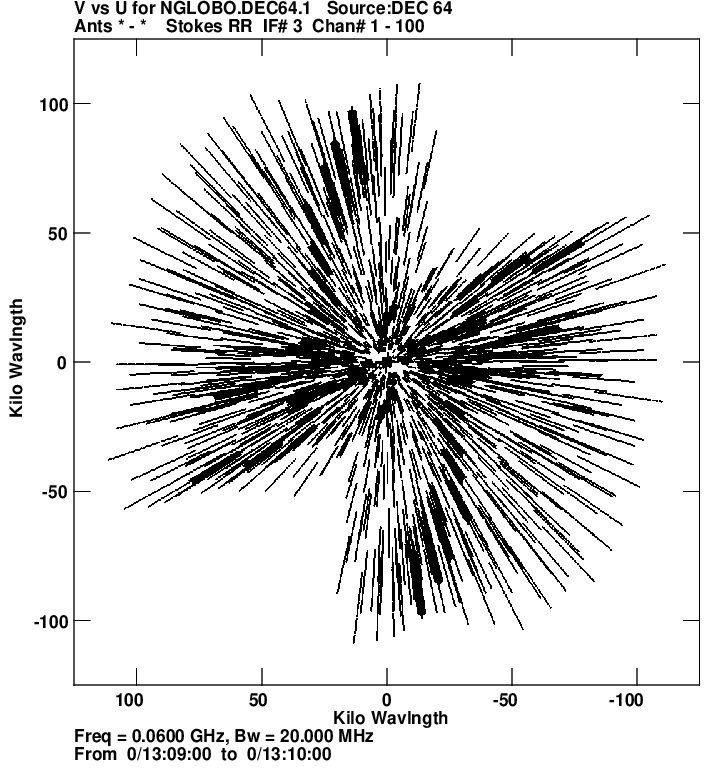}
\includegraphics[width=3.0in,angle=0]{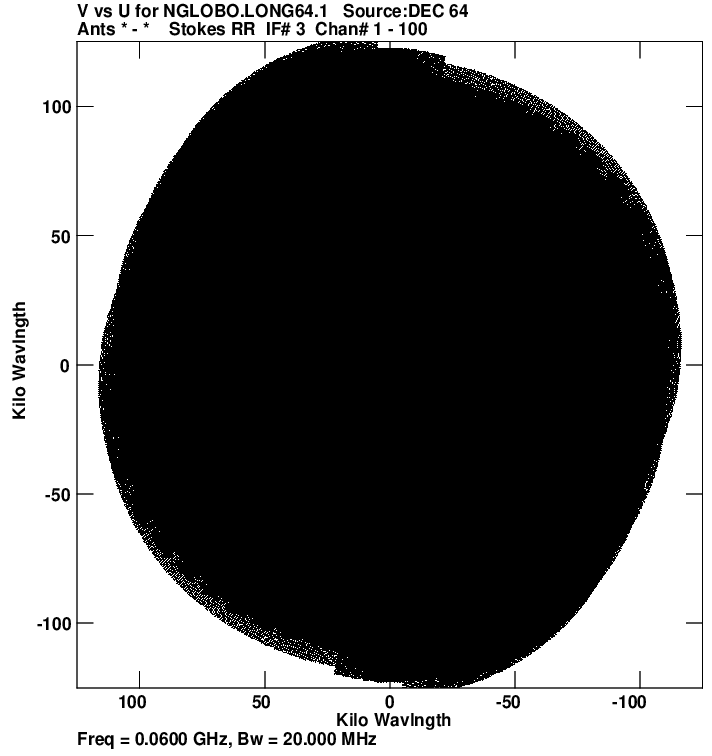}

\end{center}
\vspace{-0.5cm}
  \caption{
The ($u,v$) coverage for a source at declination +64 observed for a single integration showing the result of bandwidth synthesis at 60 MHz with a bandwidth of 20 MHz (left).  At right is the same source, frequency, and bandwidth observed over an 8 hour synthesis.
}
\label{snap20}
\end{figure}

Including bandwidth synthesis for a 20 MHz beam at 60 MHz improves the ($u,v$) coverage considerably as shown in Fig.~\ref{snap20}, and a full 8 hour track with bandwidth synthesis yields extremely good coverage.

\section{Summary: High Angular Resolution Imaging with ngLOBO}
\vspace{-0.25cm}

A renaissance in low frequencies is catalyzing an exciting new community around the wavelengths where our field was first discovered. There are two key themes behind this revolution. One is driven by Cosmology, in the search for high redshift Hydrogen from both the Epoch of Reionization and the Dark Ages. The second is motivated by the access to the high angular resolution and sensitivity catalyzed by breakthroughs in ionospheric calibration, lifting the short baseline limit that left the field abandoned for decades. The subject of this White Paper is the second theme, coupled with the opportunity afforded the development of a relatively inexpensive sub-GHz option associated with the infrastructure of the ngVLA. We call this instrument ngLOBO, as the hoped for last step in a succession of instruments currently under development, starting with LWA and VLITE and now approaching LOBO. 

A key lesson of the past decade is the unavoidable technical tension imposed by trying to achieve both key goals of low frequency radio astronomy with a single instrument. We dissipate that tension by striving for a single goal: high angular resolution imaging requiring a distributed array, in contrast to the condensed footprint driven by cosmology. Based on our own experience, we also recognize that a line must be drawn between aperture arrays embedded in the ngVLA infrastructure below 150 MHz, and a second component utilizing the ngVLA dishes above it. Recognizing this instrument cannot be perfect, we rely on the powerful approach of a non-interfering {\it commensal} system as a proven operational model that can yield over 6000 hours on the sky annually and provide a wide-field, low frequency image around every ngVLA target.

ngLOBO has three primary scientific missions: (1) Radio Large Synoptic Survey Telescope (Radio-LSST): one naturally wide beam, commensal with ngVLA, will conduct a continuous synoptic survey of large swaths of the sky for both slow and fast transients; (2) This same commensal beam will provide complementary low frequency images of all ngVLA targets and their environment {\it when such data enhances their value}. (3) Independent beams from the ngLOBO-Low aperture array will conduct research in astrophysics, Earth science and space weather applications, engaging new communities and attracting independent resources. If ngVLA operates down to 2 GHz or lower, ngLOBO data will enhance ngVLA calibration and dynamic scheduling. Non-variable field sources outside the ngVLA field of view can be harvested for serendipitous science, e.g. population studies for thermal and non-thermal continuum sources.

Finally, we have and will continue to rely on a follow-your-nose philosophy. We have spent the past decade, and longer, painstakingly developing a succession of high angular resolution capabilities based on both standalone aperture array and commensal, parabolic dish based technology. We know that our proposed approach for a much larger instrument will work because it already has on a smaller scale. A commensal ngLOBO will broad-band the ngVLA over the entire range of the radio spectrum below ALMA, and grant its user community direct, efficient, and simultaneous access in parallel to all regular ngVLA observations.

\vfill
\eject

%\tododetail{Sketch out resolution and surface brightness sensitivity for a couple
%expected configurations.  Clearly low frequencies won’t be driving
%this but we want to show that we’ve thought about it.}

%\tododetail{Maybe rehash the numbers of LWA stations/ngLOBO-High antennas we really need.}

\section{APPENDIX: Cosmology through High-redshift, Low Frequency Spectroscopy}

%\tododetail{Here in the appendix we find old LWA proposal material and other text that we can either use in the main body of the paper, or eventually just drop.}

WG4 discusses high-z molecular spectral lines accessible to ngVLA as cosmological tools. ngLOBO-High similarly has access to high-Z atomic lines, including HI 21cm absorption, OH absorption, OH megamaser emission,  and radio recombination emission lines in clean spectral windows. Depending on the station layout there is a possibility for radio recombination lines in the low band as well. The spectral resolution needed in all cases is fairly high ($\Delta v\sim 10$ km s$^{-1}$, corresponding to $\Delta \nu\sim 10$ kHz at an observational frequency of 300 MHz), and will not be possible with standard continuum observations.

\subsection{HI Absorption}

HI 21cm emission studies provide a wealth of information about the neutral ISM in local galaxies, such as gas temperature, distribution and kinematics. Emission studies at high redshift are not yet possible due to the weakness of the 21cm line (the highest redshift for HI 21cm emission detection is $z\sim0.26$, Catinella et al. 2008).    As a result, 21cm absorption in the radio and Lyman$-\alpha$ and metal-line absorption in the optical and UV remain the primary sources of information about high-redshift galaxies.   

HI 21cm absorbers primarily fall into two classes: associated and intervening. Associated systems lie at roughly the same redshift as the radio source, and the absorption arises in the neutral gas of the host galaxy. Intervening systems are galaxies which are at a lower redshift on the sightline to the radio source. Intervening galaxies, like the analogous damped Lyman$-\alpha$ (DLA) galaxy population, are considered the high redshift counterparts of normal galaxies. Information on their temperatures, metallicities, sizes, and kinematics are all important to understanding galaxy evolution. 

ngLOBO-High, with its planned broad frequency coverage from $0.05 < \nu < 1.8$ GHz, offers a unique database for a blind search for both 21cm emission at its upper frequency end ($\nu > 1.2$ GHz, $z < 0.2$), and high redshift 21cm absorption from the line rest frequency down to the lower frequency cutoff near .15~GHz ($0 < z < 8.5$).   

Experience with VLITE indicates that a commensal instrument will at PBand will view roughly $15\%$ of the visible sky in one year with 1 hour integration, and $5\%$ at 5 hours.  We use these as our target integration lengths for absorption.  The field of view, and thus the sky coverage will decrease with increasing frequency, so the actual fraction of the sky over which we can search for spectral lines increases at higher redshifts.  We use the predicted instrument sensitivity listed in Table 2, and assume that it is higher by a factor of $1.5\times$ at frequencies $700 < \nu < 1000$ MHz, and by a factor of $2\times$ at frequencies $\nu > 1000$ MHz.

The typical emission line profile has a width of $200$ km s$^{-1}$ (Papastergis et al. 2011).  With a velocity resolution of $20$ km s$^{-1}$, we would be able to detect a galaxy with $M_{HI} \sim 1 \times 10^{10}$M$_{sun}$ at $5\sigma$ in 5 hours integration at redshifts $z < 0.15$.  For fields with deeper integrations we would be able to push this limit to either a higher redshift or a lower mass upper limit.  Lensed systems at redshifts $z \sim 0.3$ should also be detectable with the ngLOBO system in a similar integration time (Hunt et al. 2016).  There are many instruments which are suitable for high redshift HI emission mapping (eg. GBT, SKA); ngLOBO-High's contributions to this area would be modest.

HI 21cm absorbers span a range of optical depths, but the highest redshift absorbers are typically only a few percent optical depth. The absorption line widths vary from very narrow ($\sim 10$~km s$^{-1}$) up to a few hundreds of km s$^{-1}$ (Kanekar et al. 2014).   For blind detection we target a $1\%$ optical depth line detected at $5\sigma$, with a width of $50$~km s$^{-1}$, spread over $\sim 5$ spectral channels.  Assuming the absorbing gas completely covers the background or associated radio source, in one hour we would be able to detect 21cm absorption against a radio source with flux density $\geq 0.75$ Jy bm$^{-1}$ at redshifts $z < 4$, and flux density $ \geq 1$ Jy bm$^{-1}$ at $z > 4$.  The broad instantaneous frequency coverage means that ngLOBO-high will be able to search the entire sightline out to the redshift of the radio source for each source in the wide field of view, excluding any regions of RFI.  Nearly all surveys for redshifted 21cm absorption target known metal-line systems such as MgII absorbers (eg. Dutta et al. 2017), Damped Lyman-$\alpha$ (DLA) systems (eg. Kanekar et al. 2013), or radio quasar samples (eg. Aditya et al. 2016). A blind radio survey for 21cm absorption on this scope would be unique and could provide useful constraints on the evolution of neutral gas in the universe without any optical bias.

\subsection{OH Megamasers}

OH Megamasers (OHMs) trace some of the most extreme physical conditions in the universe. The maser emission can be used to measure extragalactic magnetic fields and gas kinematics, and signals extreme star formation, and possibly merging black holes. The total number of detected OHMs, however, remains low. No OHMs at a distance of greater than 1300 Mpc (z = 0.265) have been detected, and no large, systematic survey for high-z OHMs has been carried out with an interferometer. The association of OHMs with bright IR merging galaxies suggests that the density of OHMs may be higher at higher redshifts, coinciding with an increase in both galaxy merger and star formation rates. It has been suggested that future searches for OHMs at high redshifts may benefit from interferometric observations, since celestial
RFI is uncorrelated from dish to dish (Willett 2012). 

ngLOBO-High's broad frequency coverage will provide an unprecedented opportunity to study OHMs from the local universe out to very high redshifts ($0 < z < 6$).  We assume that the nominal ngLOBO sensitivity is constant at $\nu < 700$ MHz, and increases by $1.5\times$ at $700 < \nu < 1000$ MHz and $2\times$ at $\nu > 1000$ MHz.  We assume an absorption line width of $\Delta v = 150$ km s$^{-1}$ spread over 5 spectral channels, and a $3\sigma$ peak.  The luminosity of the OH line is dependent on the luminosity distance, which we take from Ned Wright's calculator with a flat universe in which H$_o = 69.6$ km s$^{-1}$ Mpc$^{-1}$, $\Omega_{M} = 0.286$ and $\Omega_{vac} = 0.714$ (Wright 2006).    

We estimate that in 1 hour that ngLOBO could detect an OHM with L$_{OH} \geq 10^{3}$~L$_{\bigodot}$ at redshifts $z < 0.25$, increasing to L$_{OH} \geq 10^{4}$~L$_{\bigodot}$ at $z = 0.75$, and L$_{OH} \geq 10^{5}$~L$_{\bigodot}$ at $z = 2.5$.   Based on sky coverage from VLITE, we estimate that the corresponding sky area at 1 hour depth in 1 year for each redshift is $\Omega_{1hr, 1yr} \sim 200, 350,$ and 1500 sq. degrees.  At higher redshifts we would be able to study OHMs on sitelines with a longer integration time; however we estimate that we could detect an OHM with L$_{OH} \geq 10^{5}$~L$_{\bigodot}$ at $z = 4.5$ in about 10 hours.  While the estimated sky coverage is only a few percent at that depth, and it is unknown if the local number counts for OHMs can be extrapolated to those redshifts, we would be able to start placing limits on these signposts for star formation in dense environments at high redshifts.

\clearpage

\def\begref{\parindent=0pt\frenchspacing\parskip=0pt
    \everypar={\hangindent=0.42in}}
\def\ref{\par \noindent}
\def\HI {H\thinspace{\sc i}}
\def\txs {TXS\thinspace2226{\tt -}184}
\def\nat {Nature}
\def\apj {ApJ}
\def\apjl {ApJL}
\def\mnras {MNRAS}
\def\grl {GRL}
\def\aap {A\&A}
\def\aapr {A\&ARv}
\def\araa {ARAA}
\def\pks {PKS\thinspace2322{\tt -}123}
\def\etal{{\it et al.}}
\def\nref{\parskip0pt\par\noindent\hangindent\parindent\hangafter1}


\begin{thebibliography}{}

\bibitem[Aditya et al.(2016)]{2016MNRAS.455.4000A} Aditya, J.~N.~H.~S., Kanekar, N., \& Kurapati, S.\ 2016, \mnras, 455, 4000 

\bibitem[Anderson \& Rudnick(1996)]{2016ApJ...456...234A} Anderson, M.~C. \& Rudnick, L. \ 1996, \apj, 456, 234

\bibitem[Arzoumanian et al.(2015)]{2015ApJ...813...65T} Arzoumanian, Z., Brazier, A., et al.\ 2015, \apj, 813, 65 


\bibitem[Ascasibar \& Markevitch(2006)]{2006ApJ...650..102A} Ascasibar, Y., \& Markevitch, M.\ 2006, \apj, 650, 102 


\bibitem[Backer et al.(1982)]{Backer82}
Backer, D. C., Kulkarni, S. R., Heiles, C., Davis, M. M., Goss, W. M. 1982, Nature, 300, 615

\bibitem[Bell et al.(2016)]{Bell2016}
Bell, M.~E. et al. 2016, MNRAS, 461, 908

\bibitem[Bhakta et al.(2017)]{bdf+17} 
Bhakta, D. et al. 2017, MNRAS, submitted.

\bibitem[Begelman et al.(1984)]{1984RvMP...56..255B} Begelman, M.~C.,
Blandford, R.~D., \& Rees, M.~J.\ 1984, Reviews of Modern Physics, 56, 255

\bibitem[Bowman et al.(2008)]{2008ApJ...676....1B} Bowman, J.~D., Rogers, A.~E.~E., \& Hewitt, J.~N.\ 2008, \apj, 676, 1-9 

\bibitem[Brogan et al.(2003)]{B+2003} Brogan et al. \ 2003, Astronomische Nachrichten Supplementary,  324, 17

\bibitem[Brogan et al.(2005a)]{2005ApJ...629L..105B} Brogan et al. \ 2005a, \apjl, 629, L105 

\bibitem[Brogan et al.(2005b)]{2005AJ...130..148B} Brogan et al. \ 2005b, AJ, 130, 148 

\bibitem[Brogan et al.(2006)]{2006ApJ...639..25B} Brogan et al. \ 2006, ApJ, 639, 25 


\bibitem[Brunetti et al.(2008)]{2008Natur.455..944B} Brunetti, G., Giacintucci, S., Cassano, R., et al.\ 2008, \nat, 455, 944 


\bibitem[Brunetti \& Jones(2014)]{2014IJMPD..2330007B} Brunetti, G., \& Jones, T.~W.\ 2014, International Journal of Modern Physics D, 23, 1430007 

\bibitem[Budden(1985)]{1985prw..book.....B} Budden, K.~G.\ 1985, The propagation of radio waves..~K.~G.~Budden.Cambridge University Press, Cambridge  

\bibitem[Cassano et al.(2010)]{2010ApJ...721L..82C} Cassano, R., Ettori, S., Giacintucci, S., et al.\ 2010, \apjl, 721, L82 

\bibitem[Cassano et al.(2013)]{2013ApJ...777..141C} Cassano, R., Ettori, S., Brunetti, G., et al.\ 2013, \apj, 777, 141 

\bibitem[Catinella et al.(2008)]{2008ApJ...685L..13C} Catinella, B., Haynes, M.~P., Giovanelli, R., Gardner, J.~P., \& Connolly, A.~J.\ 2008, \apjl, 685, L13 

\bibitem[Chatterjee et al.(2017)]{chatt+17} Chatterjee, S., Law, C.~J. et al.\ 2005, \nat, 434, 50

\bibitem[Chowla et al.(2017)]{2017} Chowla, P. et al. 2017, \apj, (https://arxiv.org/abs/1701.07457, accepted).

\bibitem{2011ursi.confE...5C} Clarke, T.~E., Perley, R.~A., Kassim, N.~E., et al.\ 2011, General Assembly and Scientific Symposium, 2011 XXXth URSI

\bibitem[Clarke et al.(2009)]{2009ApJ...697.1481C} Clarke, T.~E., Blanton, E.~L., Sarazin, C.~L., et al.\ 2009, \apj, 697, 1481 

\bibitem{clarke14}Clarke, T.E., Higgins, C., Skarda, J., Imai, K., Imai, M., Reyes, F., Thieman, J., Jaeger, T., Schmitt, H., Dalal, N.P., Dowell, J., Ellingson, S.W., Hicks, B., Schinzel, F.K, \& Taylor, G.B. 2014, JGR, 119, 9508 

\bibitem[Condon et al.(1998)]{1998AJ....115.1693C} Condon, J.~J., Cotton, W.~D., Greisen, E.~W., et al.\ 1998, AJ, 115, 1693 

\bibitem{crane10}Crane, P. 2010, LWA Memo series, Number 168

\bibitem{cran17}Cranmer, M.D., Barsdell, B.R.,  et al. 2017, submitted to JAI, 
astro-ph/1708.00720

\bibitem[Cuciti et al.(2015)]{2015A&A...580A..97C} Cuciti, V., Cassano, R., Brunetti, G., et al.\ 2015, \aap, 580, A97 


\bibitem[de Gasperin et al.(2014)]{2014MNRAS.444.3130D} de Gasperin, F., van Weeren, R.~J., Br{\"u}ggen, M., et al.\ 2014, \mnras, 444, 3130 


\bibitem[Delaney et al.(2015)]{delaney2015} Delaney, T., Kassim, N.~E. et al. \ 2015, \apj, 785, 7 

\bibitem[Erickson \& Mahoney(1985)]{EM2015} Erickson, W.~C. \& Mahoney, M.~J. \ 1985, \apj, 299, 29 

\bibitem[Dutta et al.(2017)]{dutta2017} Dutta, R., Srianand, R., Gupta, N., \& Joshi, R.\ 2017, \mnras, 468, 1029 


\bibitem[Fabian(1994)]{1994ARAA..32..277F} Fabian, A.~C.\ 1994, \araa, 32, 277 


\bibitem[Fabian et al.(2002)]{2002MNRAS.331..369F} Fabian, A.~C., Celotti, A., Blundell, K.~M., Kassim, N.~E., \& Perley, R.~A.\ 2002, \mnras, 331, 369 

\bibitem[Fabian et al.(2006)]{2006MNRAS.366..417F} Fabian, A.~C., Sanders, J.~S., Taylor, G.~B., et al.\ 2006, \mnras, 366, 417 

\bibitem[Fabian(2012)]{2012ARAA..50..455F} Fabian, A.~C.\ 2012, \araa, 50, 455 


\bibitem[Feretti et al.(2012)]{2012AARv..20...54F} Feretti, L., Giovannini, G., Govoni, F., Murgia, M.\ 2012, \aapr, 20, 54 

\bibitem[Gardner \& Liu(2010)]{gar10} Gardner, C.S., \& Liu, A.Z.
2010, JGR Atmospheres, 115, D20302

\bibitem[de Gasperin et al.(2012)]{2012AA...547A..56D} de Gasperin, F., Orr{\'u}, E., Murgia, M., et al.\ 2012, A\&A, 547, A56

\bibitem[Giacintucci et al.(2012)]{2012ApJ...755..172G} Giacintucci, S., O'Sullivan, E., Clarke, T.~E., et al.\ 2012, \apj, 755, 172 

\bibitem[Giacintucci et al.(2014)]{2014ApJ...781....9G} Giacintucci, S., Markevitch, M., Venturi, T., et al.\ 2014, \apj, 781, 9 

\bibitem[Giacintucci et al.(2017)]{2017ApJ...841...71G} Giacintucci, S., Markevitch, M., Cassano, R., et al.\ 2017, \apj, 841, 71 

\bibitem[Gitti et al.(2011)]{2011ApJ...732...13G} Gitti, M., Nulsen,
P.~E.~J., David, L.~P., McNamara, B.~R.,
\& Wise, M.~W.\ 2011, ApJ, 732, 13

\bibitem[van Haarlem et al.(2013)]{2013AA...556A...2V}
van Haarlem, M.~P., Wise, M.~W., Gunst, A.~W., et al.\ 2013, A\&A, 556, A2

\bibitem[Helmboldt et al.(2012)]{hel12a}{Helmboldt}, J.~F., W.~M. {Lane}, and W.~D. {Cotton}, 2012, Radio Science, 47, RS5008

\bibitem[Helmboldt and Intema (2012)]{hel12b}{Helmboldt}, J.~F., and H.~T. {Intema}, 2012,
  Radio Science, 47, RS0K03

\bibitem[Helmboldt et al.(2013)]{hel13}Helmboldt, J.F., Clarke, T.E., Craig, J., Ellingson, S.W., Hartman, J.M., Hicks, B.C., Kassim, N.E., Taylor, G.B., \& Wolfe, C.N. 2013, Radio Science, 48, 491 

\bibitem{hel14}
Helmboldt, J.F., Ellingson, S.W., Taylor, G.B., Wilson, T.L., \& Wolfe, C.N.
2014, Radio Science, 49, 3

\bibitem[Helmboldt et al. (2015)]{hel15} Helmboldt, J.F., N.E.\ Kassim, S.W.\ Teare, 2015, Earth and Space Science, 2, 387--402

\bibitem[Helmboldt (2016)]{hel16}Helmboldt, J.F. 2016, Ann Geophys, 34, 529 

\bibitem[Hessels et al.(2014]{hess14} Hessels, J.~W.~T. et al. 2014, ASI Conference Series, 2014,13, 43

\bibitem[Hitomi Collaboration et al.(2016)]{2016Natur.535..117H} Hitomi Collaboration, Aharonian, F., Akamatsu, H., et al.\ 2016, \nat, 535, 117 


\bibitem{how11} Howard, T.A., Coronal Mass Ejections, An Introduction, Springer, New York, doi: 10.1007/978-
1-4419-8789-1, 2011

\bibitem[Howard et al.(2016)]{2016ApJ...831..208H} Howard, T.~A., Stovall, K., Dowell, J., Taylor, G.~B., \& White, S.~M.\ 2016, \apj, 831, 208 

\bibitem[Hunt et al.(2016)]{2016AJ....152...30H} Hunt, L.~R., Pisano, D.~J., \& Edel, S.\ 2016, Astronomical Journal, 152, 30 

\bibitem[Hurley-Walker et al.(2017)]{2017MNRAS.464.1146H} Hurley-Walker, N., Callingham, J.~R., Hancock, P.~J., et al.\ 2017, \mnras, 464, 1146 

\bibitem[Hyman et al.(2005)]{hlk+05} Hyman, S.~D., Lazio, T.~J.~W., Kassim, N.~E., et al.\ 2005, \nat, 434, 50

\bibitem[Intema et al.(2017)]{2017AA...598A..78I} Intema, H.~T., Jagannathan, P., Mooley, K.~P., \& Frail, D.~A.\ 2017, \aap, 598, A78 

\bibitem[Jacobson and Erickson (1992)]{jac92} {Jacobson}, A.~R., and W.~C. {Erickson}, 1992, Planetary and Space Science, 40,
  447--455

\bibitem{jak15}
Jakosky, B.~M. et al. 2016, GRL, 42, 8791

\bibitem[Kale et al.(2015)]{2015AA...579A..92K} Kale, R., Venturi, T., Giacintucci, S., et al.\ 2015, \aap, 579, A92 

\bibitem[Kanekar et al.(2013)]{2013MNRAS.428..532K} Kanekar, N., Ellison, S.~L., Momjian, E., York, B.~A., \& Pettini, M.\ 2013, \mnras, 428, 532 

\bibitem[Kanekar et al.(2014)]{2014MNRAS.438.2131K} Kanekar, N., Prochaska, J.~X., Smette, A., et al.\ 2014, \mnras, 438, 2131 

\bibitem{1989ApJ..347..915K} Kassim, N.~E. et al.\ 1989, \apj, 347, 915 

\bibitem{2007ApJS..172..686K} Kassim, N.~E., Lazio, T.~J.~W., Erickson, W.~C., et al.\ 2007, ApJS, 172, 686 

\bibitem{1989ApJ..338..152} Kassim, N.~E. et al.\ 1989, ApJ, 338, 152 

\bibitem{koc14}Kocz, J., Greenhill, L.J., Barsdell, B.R., Price, D., Bernardi, G., Bourke, S., Clark, M.A., Craig, J., Dexter, M., Dowell, J., Eftekhari, T., Ellingson, S., Hallinan, G., Hartman, J., Jameson, A., MacMahon, D., Taylor, G.B., Schinzel, F., \& Werthimer, D. 2014, JAI, 450003 

\bibitem[Kramer \& Champion(2013)]{2013CQGra..30v4009K} Kramer, M., \& Champion, D.~J.\ 2013, Classical and Quantum Gravity, 30, 224009 

\bibitem[Lane et al.(2014)]{2014MNRAS.440..327L} Lane, W.~M., Cotton, W.~D., van Velzen, S., et al.\ 2014, \mnras, 440, 327

\bibitem[Lang, C.C. et al.(1999)]{1999ApJL...521, 41L} Lang, C.~C. et al.\ 1999, \apj, 521, 41L

\bibitem{2000AJ..119..207L} LaRosa, T.~N., Kassim, N.~E. et al.\ 2000, Astronomical Journal, 119, 207 

\bibitem[Lazio et al.(2004)]{2004ApJ...612..511L} Lazio, T.~J., W., Farrell, W.\
~M., Dietrick, J., et al.\ 2004, \apj, 612, 511

\bibitem[Loi et al.(2015)]{2015GeoRL..42..3707L} Loi, S,~T. et al. \ 2015, Geophysical Research Letters, 42, L3707

\bibitem[Longair 1990] {1990Springer.262...227} Longair, M.~S., in Low Frequency Astrophysics from Space., eds.~N.~E.~Kassim \& K.~W. Weiler, Springer-Verlag Lecture Notes in Physics, 262, 227   

\bibitem[Lynch et al.(2017)]{lyn17} Lynch, C.~R., Lenc, E., Kaplan, D.~L., Murphy, T., Anderson, G.~E. 2017, \apjl, 836, L30

\bibitem[Mahoney \& Erickson(1985)]{ME2015} Mahoney, M.~J. \& Erickson, W.~C. \ 1985, \nat, 317, 154 


\bibitem[Marr et al.(2014)]{2014ApJ...780..178M} Marr, J.~M., Perry, T.~M.,
Read, J., Taylor, G.~B., \& Morris, A.~O.\ 2014, ApJ, 780, 178

\bibitem[Markevitch \& Vikhlinin(2007)]{2007PhR...443....1M} Markevitch, M., \& Vikhlinin, A.\ 2007, physrep, 443, 1 


\bibitem[Mittal et al.(2009)]{2009AA...501..835M} Mittal, R., Hudson, D.~S., Reiprich, T.~H., \& Clarke, T.\ 2009, \aap, 501, 835

\bibitem[McNamara et al.(2000)]{2000ApJ...534L.135M} McNamara, B.~R., Wise,
M., Nulsen, P.~E.~J., et al.\ 2000, ApJL, 534, L135

\bibitem[Monsalve et al.(2017)]{2017ApJ...835...49M} Monsalve, R.~A., Rogers, A.~E.~E., Bowman, J.~D., \& Mozdzen, T.~J.\ 2017, \apj, 835, 49 

\bibitem[Murgia et al.(2011)]{2011A&A...526A.148M} Murgia, M., Parma, P., Mack, K.-H., et al.\ 2011, \aap, 526, A148 


\bibitem[Murphy et al.(2015)]{2015MNRAS.446.2560M} Murphy, T., Bell, M.~E., Kap\
lan, D.~L., et al.\ 2015, MNRAS, 446, 2560

\bibitem[Murphy et al.(2017a)]{2017MNRAS.466.1944M} Murphy, T. et al.\ 2017, MNRAS, 466, 1944

\bibitem[Murphy et al.(2017b)]{2017PASA.34.20M} Murphy, T., et al.\ 2017, PASA, 34, 20

\bibitem[Nord et al.(2006)]{2006AJ...132..242} Nord et al. \ 2006, AJ, 132, 242 

\bibitem{kso14b} Obenberger, K.S., Taylor, G.B., Hartman, J.M. et al. 2014, ApJL, 788, L26

\bibitem[Obenberger et al.(2015)]{2015JGRA..120.9916O} Obenberger, K.~S., Taylor, G.~B., Lin, C.~S., et al.\ 2015, Journal of Geophysical Research (Space Physics), 120, 9916 

\bibitem[Obenberger et al.(2016)]{2016GeoRL..43.8885O} Obenberger, K.~S., Holmes, J.~M., Dowell, J.~D., et al.\ 2016, \grl, 43, 8885 

\bibitem[Papastergis et al.(2011)]{2011ApJ...739...38P} Papastergis, E., Martin, A.~M., Giovanelli, R., \& Haynes, M.~P.\ 2011, \apj, 739, 38 

\bibitem[Peterson \& Fabian(2006)]{2006PhR..427..1P} Peterson, J.~R., \& Fabian, A.~C.\ 2006, physrep, 427, 1 

\bibitem[Polisensky et al.(2011)]{2016ApJ...832...60P} Polisensky, E., Lane, W.~M., Hyman, S.~H., et al.\ 2016, \apj, 832, 60 

\bibitem[Rajwade \& Lorimer(2017)]{2016MNRAS.465.2286R} Rajwade, K.~M. \& Lorimer, D.~R.\ 2017, \mnras, 465, 2286

\bibitem[Reardon et al.(2016)]{2016MNRAS.455.1751R} Reardon, D.~J., Hobbs, G., Coles, W., et al.\ 2016, \mnras, 455, 1751 

\bibitem[Rengelink et al.(1997)]{1997AAS..124..259R} Rengelink, R.~B., Tang, Y., de Bruyn, A.~G., et al.\ 1997, A\&AS, 124, 259  

\bibitem[Reynolds \& Ellison(1992)]{1992...399..75R} Reynolds, S.~P. \& Ellison, D.~C.\ 1992, \apjl, 399, 75 

\bibitem[Rieger et al.(2007)]{2007ApSS.309..119R} Rieger, F.~M., Bosch-Ramon, V., \&
Duffy, P.\ 2007, Ap\&SS, 309, 119

\bibitem[Savini et al.2015] Savini, F., Thesis, Universita di Pisa

\bibitem[Schutz \& Ma(2016)]{2016MNRAS.459.1737S} Schutz, K., \& Ma, C.-P.\ 2016, \mnras, 459, 1737 


\bibitem[Stappers et al.(2011)]{2011AA...530..80S} Stappers, B.~W. et al.\ 2011, A\&A, 530, 80 



\bibitem[Stovall et al.(2015)]{2015ApJ...808..156S} Stovall, K., Ray, P.~S., Blythe, J., et al.\ 2015, ApJ, 808, 156 

\bibitem{tay12} Taylor, G.~B., Ellingson, S.~W., Kassim, N.~E., et al.\ 2012, Journal of Astronomical Instrumentation, 1, 50004 

\bibitem[van Weeren et al.(2011)]{2011AA...533A..35V} van Weeren, R.~J., Br{\"u}ggen, M., R{\"o}ttgering, H.~J.~A., et al.\ 2011, \aap, 533, A35 


\bibitem[Webber 1990] {1990Springer.262...217} Webber, W.~R., in Low Frequency Astrophysics from Space., eds.~N.~E.~Kassim \& K.~W. Weiler, Springer-Verlag Lecture Notes in Physics, 262, 217   

\bibitem[Winglee et al.(1986)]{1986ApJ...309L..59W} Winglee, R.~M., Dulk, G.~A.\
, \& Bastian, T.~S.\ 1986, ApJL, 309, L59

\bibitem[Yuan et al.(2015)]{2015ApJ...813...77Y} Yuan, Z.~S., Han, J.~L., \& Wen, Z.~L.\ 2015, \apj, 813, 77

\bibitem[Zhuravleva et al.(2014)]{2014Natur.515...85Z} Zhuravleva, I., Churazov, E., Schekochihin, A.~A., et al.\ 2014, \nat, 515, 85 

\bibitem[ZuHone et al.(2013)]{2013ApJ...762...78Z} ZuHone, J.~A., Markevitch, M., Brunetti, G., \& Giacintucci, S.\ 2013, \apj, 762, 78 



\end{thebibliography}
\end{document}